\documentclass[5p, twocolumn,authoryear]{elsarticle}

\usepackage{amsmath,amssymb}
\usepackage{graphicx}
\usepackage{booktabs}
\usepackage{siunitx}
\usepackage{hyperref}
\usepackage{xcolor}
\usepackage{multirow}
\usepackage{dashrule}
\usepackage{subcaption}

\begin{document}

\begin{frontmatter}

\title{Semi supervised GAN for smart microscopy, fast and data efficient cell cycle classification}

\author[inst1]{Rajeev Manick}
\author[inst1]{Youssef El Habouz$^{*}$}
\author[inst1]{Ma\"elle Guillout}
\author[inst2]{Celia Martin}
\author[inst1]{Julia Bonnet}
\author[inst1]{Louis Ruel}
\author[inst1]{Sylvain Pastezeur}
\author[inst2]{Olivier Chanteux}
\author[inst2]{Otmane Bouchareb}
\author[inst1,inst3]{Marc Tramier}
\author[inst1]{Jacques P\'ecr\'eaux}
\cortext[$^{*}$]{Co-first author.}
\cortext[]{Corresponding authors: rajeev.manick@univ-rennes.fr, \\jacques.pecreaux@univ-rennes.fr}



\affiliation[inst1]{organization={CNRS, Univ. Rennes, IGDR},
            addressline={UMR 6290},
            city={F-35043 Rennes},
            country={France}}
\affiliation[inst2]{organization={Inscoper SAS},
            addressline={F-35510},
            city={Cesson-S\'evign\'e},
            country={France}}
\affiliation[inst3]{organization={Univ. Rennes, BIOSIT},
            addressline={UMS CNRS 3840, US INSERM 018},
            city={F-35000 Rennes},
            country={France}}


%
%
%
%
%
%
%
%
%
%
%

\begin{abstract}
Modern optical microscopes are fully motorised; however, transforming them into truly smart systems requires real-time adjustment of acquisition settings in response to detected objects and dynamic biological events. At the core are classification algorithms that commonly depend on customised softwares and are generally designed for narrowly-defined biological applications. In addition, they often require substantial annotated datasets for effective training.
We introduce a semi-supervised generative adversarial network (SGAN) for robust cell-cycle stage classification under low-resource conditions, adaptable to diverse cellular structures. The framework combines unlabelled microscopy images with synthetically generated samples to mitigate limited annotation, while preserving stable performance even when the unlabelled subset is class-imbalanced. Tested on the Mitocheck dataset, which features five mitosis classes, the model achieved $93 \pm 2\%$ accuracy using only 80 labelled per class and 600 unlabelled images. The proposed algorithm is generic and can be readily adapted to new labeling schemes, classification targets, cell lines, or microscopy modalities through transfer learning. SGAN is well suited for integration into automated microscopes, enabling efficient and adaptable image analysis across diverse biological and microscopy applications.
\end{abstract}


\end{frontmatter}


\section{Introduction}

Through its diverse modalities, optical microscopy enables unparalleled approaches to investigate the living. Beyond gaining an in-depth understanding of cell biology, it is widely used in high-content screening (HCS) to test new drugs \citep{balasubramanian23,daetwyler23,renz13}. The progress in biology is now tightly linked to progress in imaging, leading to a vast range of modalities beyond the bare observation \citep{daetwyler23,lelek21,mangeat21,prakash22}. Fluorescence microscopy can also capture functional aspects, such as the dynamics of labelled proteins, through high-frame-rate movies and so-called F-techniques, which are based primarily on fluorescence recovery after photobleaching (FRAP), photoconversion, and fluorescence correlation spectroscopy (FCS, \citet{ishikawaankerhold12}. Such techniques also enable access to interactions between proteins, e.g., using Förster Resonance Energy Transfer (FRET) or fluorescence lifetime imaging microscopy (FLIM), and even allow for the perturbation of the sample through light to observe its adapt \citep{demedeiros20,khamo17,shakoor22}. While tremendous progress in optics and electronics supported this evolution, these experiments often remain tour-de-force, requiring dual expertise in biology to identify events and objects of interest, and in microscopy to tweak ever more complex imaging systems. 

To address this limitation, smart microscopy emerged, aiming at condensing some expertise in the accompanying software, which partly stands in for the experimenter \citep{daetwyler23,eisenstein20, eisenstein23, hinderling26,scherf15}. Such software is tasked with finding the, often rare, objects of interest by an automaton in the so-called generic event-driven acquisition \citep{almada19, andre23, bonnet24, conrad11, durand18, fox22,mahecic22, royer16, shi24, stepp25}. Most smart microscopes are used to grab a first round of images, a so-called screening sequence, and analyse them to find objects of interest, then to instruct the microscope to grab these objects in further detail with different imaging modalities \citep{carro15, conrad11, meng22, stepp25}. The analysis could be performed by the computer gathering the images or by a dedicated device inserted between this computer and the camera. In all cases, it runs an image processing workflow to identify and localise the objects of interest. To capture very transient and dynamic events, we set out to analyse images from the screening sequence on the fly and interrupt it when finding an object or event of interest \citep{balluet22, bonnet24}.

Pivotal to smart microscopes is the algorithm analysing the images. The first implementations aimed at using on-the-fly analysis to improve the quality of a specific experiment or tune one of the devices attached to the microscope. For instance, to improve acquisition, it enables adjusting automatically adaptive optics to correct aberrations \citep{hu23}, performing an acquisition only at the place or time of interest, saving sample illumination and thus photobleaching, phototoxicity and time  \citep{abouakil21, lang12, wenus09}, or drive laser micro-dissection \citep{meng22}. Alternatively, it may also directly take part in the experiment, for instance, by analysing live cell fluorescence as a readout of intracellular activity and either (i) drive a microfluidic device to subject cells to chemicals promoting or repressing specific gene expression \citep{lugagne17, perrino19}, or (ii) control light that in turn control gene expression through optogenetic \citep{chait17, toettcher11}. However, a common trait of these approaches is the hard-coded analysis of the images. Modern deep Learning approaches have been a game changer and offer a highly promising alternative for enabling users to design a wide range of experiments.

To ensure fast-enough detection of objects or events of interest, phenotype classification by deep learning network proved to be highly efficient \citep{kensert19, krentzel23, nguyen21, yao19, zaritsky21}. When it comes to embedding classification into a microscope automation, most designs were specific to microscopes/devices, both in light and electron microscopy \citep{bouvette22, hermann11, mahecic22, shi23}.  Alternatively, one can let the user code the automaton that links the image processing and microscope driving using established frameworks \citep{fox22, pinkard16, pinkard21}. 

A standard limitation of microscopy imaging when it comes to deep learning is the small size of the available training dataset \citep{chai23, dou23, shaikhina17}. Beyond using a domain transfer from a pre-trained model, such a challenge may be addressed by reducing the burden of annotating using non-labelled images, including semi-supervised, unsupervised, or self-supervised approaches \citep{lu19, moen19, nguyen21, vanvalen16, wu22, yao19}. 

This work is motivated by the practical constraints of many AI-assisted real-time microscopy applications, where classifiers must learn from very limited annotation while remaining fast and reliable enough to support adaptive acquisition and autonomous decision-making. We therefore investigate a semi-supervised generative adversarial network (SGAN) as a data-efficient framework for microscopy-based cell-phase classification. Specifically, we assess the behaviour of semi-supervised learning (SSL) across a range of labelled and unlabelled data budgets, compare it with competing supervised and semi-supervised methods, and examine its potential for integration into smart microscopy workflows requiring robust, transferable, and computationally efficient image analysis.

\section{Related works}

SSL methods have been extensively studied in the literature in recent years. A comprehensive review by \citet{vanengelen20} categorizes SSL methodologies into \emph{inductive} methods, which aim to build classifiers that generalise to unseen data, and \emph{transductive} methods, which directly optimize predictions over a fixed unlabelled dataset. Inductive strategies differ mainly in how they exploit unlabelled data, including approaches based on iterative pseudo-label generation, unsupervised or self-supervised feature learning, and methods that integrate unlabelled samples explicitly into the training objective. 

Building upon this inductive perspective, our proposed approach adopts a generative adversarial network (GAN) formulation. Accordingly, this section focuses on seminal and recent GAN-based SSL models developed across various application domains, with emphasis on architectural mechanisms that incorporate unlabelled data directly into representation learning. Broader surveys of alternative SSL paradigms can be found in a review by \citet{sajun22}. 

Fundamentally, SSL relies on the assumption that the geometric structure of the data manifold encodes information about underlying class distributions. In this context, GANs are particularly well suited due to their capacity to model complex, high-dimensional data distributions, thereby enabling effective feature learning and classification in regimes where annotated data are scarce.

One of the earliest works applying GANs to SSL was the Semi-Supervised GAN proposed by \citet{odena16}. Their approach modifies the discriminator into an $N+1$-class classifier that jointly performs discrimination and class prediction, allowing unlabelled data to directly inform feature learning. Experiments by \citet{ouriha24} demonstrated that SSL approach significantly outperforms conventional CNN classifiers in low-label settings, while its advantage diminishes as supervision increases, with performance converging to standard supervised CNN baselines. These results highlighted a particularly data-efficient strategy in label-scarce environments, motivating its adoption in subsequent studies.

Later work extended semi- and self-supervised GANs from classification to large-scale image generation. \citet{lucic19} introduced S$^2$GAN and S$^3$GAN, combining self-supervised rotation prediction with semi-supervised label learning to produce high-quality conditional images. Evaluated on ImageNet ((1.3 million images, 1,000 classes at $128 \times 128$ resolution), their approach demonstrated strong label efficiency: the unsupervised variant achieved state-of-the-art generation quality, while S$^3$GAN matched and even surpassed fully supervised BigGAN \citep{brock18} using only a small fraction ($\sim$10--20\%) of labelled data.

Recent efforts of using GAN-based SSL in the classification domain has been the \emph{Triple-GAN} framework proposed by \citet{li17a}, which formulates learning as a three-player minimax game between a generator, a classifier, and a discriminator, thereby decoupling classification and discrimination objectives. Evaluated on datasets including CIFAR-10 \citep{krizhevsky09}, SVHN \citep{netzer11}, Tiny ImageNet \citep{deng09}, and STL-10 \citep{coates11} with only a small fraction of labelled data (typically 250 to 4,000 labelled images), Triple-GAN substantially reduced classification error relative to prior SSL techniques and strong baselines such as Mean Teacher \citep{tarvainen18}; for example, on CIFAR-10 with 4,000 labels, the error decreased from approximately $16$ to $18\%$ to $\approx 12\%$ without data augmentation and to $\approx 10\%$ with augmentation, while maintaining competitive conditional image generation quality.

Complementary work in the medical field by \citet{haque21} introduced EC-GAN for low-sample supervised classification. On SVHN (73,257 training images), experiments using only 10–30\% of annotated data ($\sim$7k--22k samples) achieved accuracies up to $\sim$94.3\%, outperforming both standard CNN classifiers and shared discriminator-classifier GAN architectures. On a pediatric chest X-ray dataset comprising 5,863 labelled images, EC-GAN reached $\sim$98\% accuracy using the full dataset and maintained strong performance in extreme low-data regimes as well.

More recently, \citet{manni25} introduced the SPARSE framework, a GAN-based semi-supervised learning approach specifically designed for extremely low labelled-data regimes. The method was evaluated on eleven datasets from the MedMNIST \citep{yang23} benchmark under few-shot settings of 5, 10, 20, and 50 labelled samples per class. In their experimental protocol, all remaining training samples were treated as unlabelled, resulting in substantially large unlabelled pools that often ranged from $10^4$ to $10^5$ images depending on the dataset.Mean per-class accuracy averaged across datasets reached 66.22\%, 70.95\%, 75.71\%, and 78.28\% for SPARSE$_{ens}$ in the 5-, 10-, 20-, and 50-shot settings, respectively, with the non-ensemble SPARSE model showing slightly lower performance.

Overall, these studies indicate that high classification performance with semi-supervised GANs, often exceeding $\sim$85--90\% is generally reported in experimental settings based on relatively large datasets comprising substantial numbers of labelled and/or unlabelled samples. These dataset scales remain one to two orders of magnitude larger than those typically encountered in real-world bio-imaging applications, where acquisition time, experimental throughput, and annotation costs severely constrain data availability.

\section{Dataset and splits}
\subsection{Mitocheck Dataset}
For consistency with our earlier work, we used the Mitocheck dataset assembled by \citet{balluet22} and derived from the original data of \citet{neumann10}. It consisted of wide-field fluorescence time-lapses of HeLa Kyoto cells, expressing chromatin Green Fluorescent Protein (GFP) marker. Images of this dataset were acquired with a 10$\times$ dry objective on an Olympus ScanR microscope.

\subsubsection{Labelled Data}

\begin{figure}[h]
\centering

\begin{minipage}{0.09\textwidth}
    \centering
    \includegraphics[width=\linewidth]{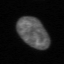}
\end{minipage}
\begin{minipage}{0.09\textwidth}
    \centering
    \includegraphics[width=\linewidth]{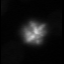}
\end{minipage}
\begin{minipage}{0.09\textwidth}
    \centering
    \includegraphics[width=\linewidth]{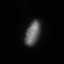}
\end{minipage}
\begin{minipage}{0.09\textwidth}
    \centering
    \includegraphics[width=\linewidth]{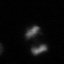}
\end{minipage}
\begin{minipage}{0.09\textwidth}
    \centering
    \includegraphics[width=\linewidth]{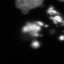}
\end{minipage}

\caption{Representative examples of the five classes used for SGAN model. 
From left to right: Interphase, Prometaphase, Metaphase, Anaphase, and Apoptosis.}

\label{figure:labeled_data}
\end{figure}

All labelled images were manually annotated by biology experts to assign ground-truth cell-cycle phase labels across five classes: Interphase, Prometaphase, Metaphase, Anaphase, and Apoptosis. Importantly, the cropped cell images were sourced from independent static microscopy snapshots rather than time-lapse sequences, ensuring that each image represents a unique cell observation with no temporal correlation between samples. We split the labelled sets with varying sizes of 20, 40, 60, and 80 images per class to systematically evaluate our model performance across different label budgets. One representative example from each class is shown in Figure~\ref{figure:labeled_data}.

A separate held-out test set of 100 images (20 labelled images per class, 
balanced across all five phases) was reserved for final model evaluation and 
was never used during training or validation. All labelled images underwent identical preprocessing: they were loaded as 8-bit grayscale, resized to 64×64 pixels, and normalized to the range [-1, 1].

\subsubsection{Unlabelled Data}
While semi-supervised learning typically assumes a regime where the number of unlabelled samples ($N_U$) is much larger than the number of labelled samples ($N_L$), i.e., $N_U \gg N_L$,
(e.g. \citealt{wang21, wang23, zhong25}, real-world biomedical imaging research often operate under severe data scarcity constraints. To reflect this practical reality and evaluate SGAN robustness in low-data regimes, we constrain the unlabelled pool to a total of $N_U = 600$ images.

\begin{figure*}
\centering
\includegraphics[width=.48\linewidth]{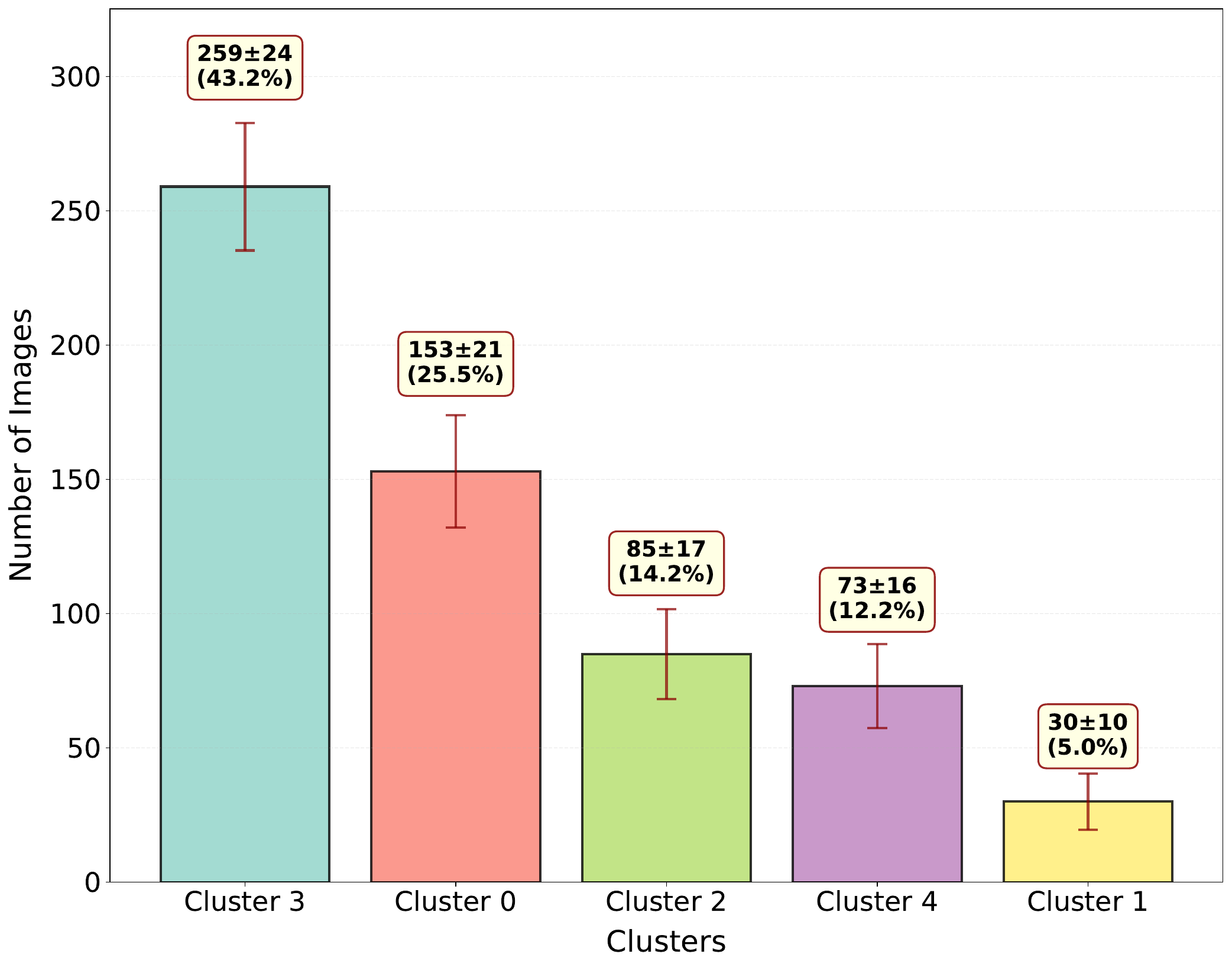} 
\includegraphics[width=.48\linewidth]{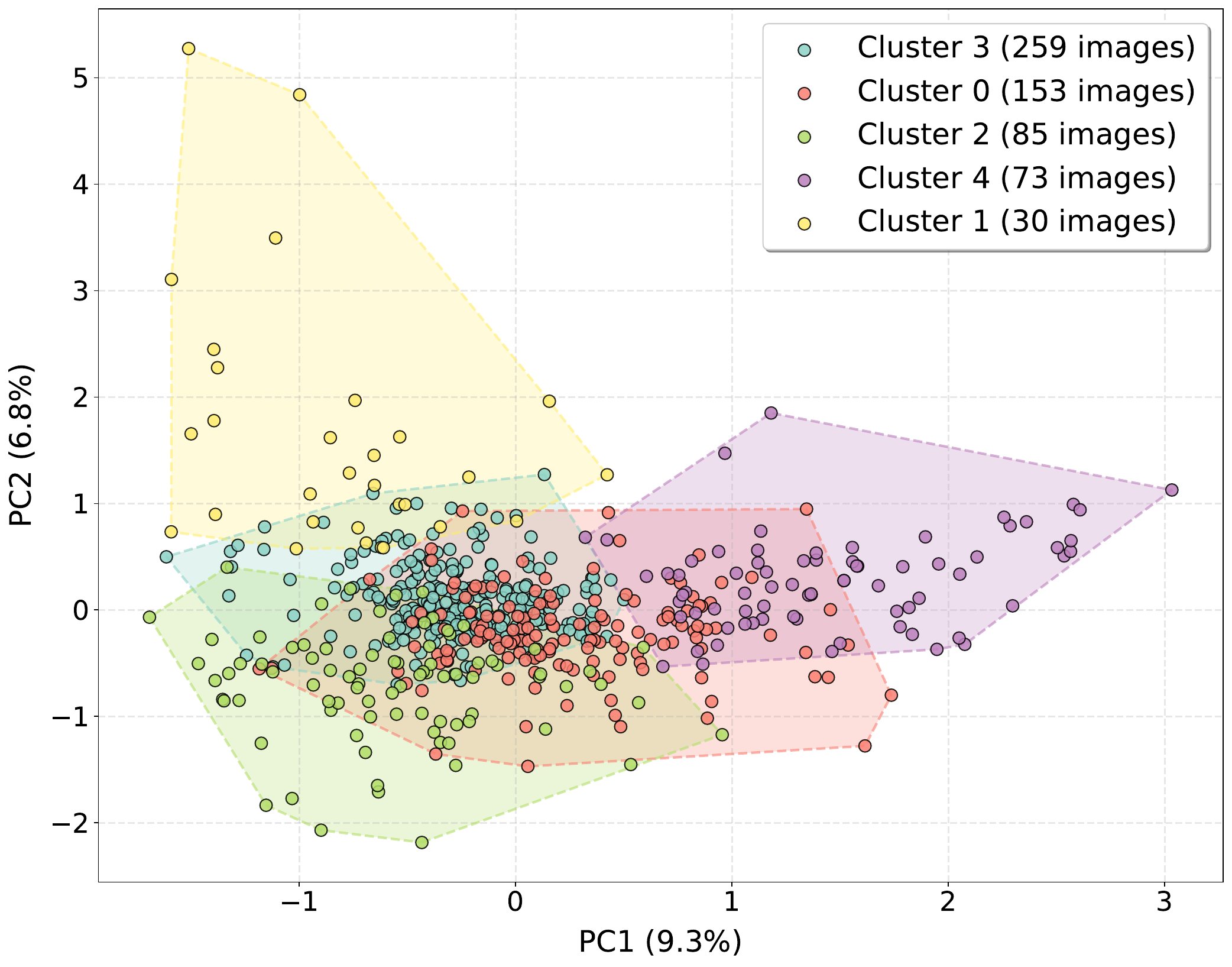} 
\caption{Unsupervised clustering analysis of 600 unlabelled cell images using Blob detection features reveals natural five-cluster decomposition with pronounced class imbalance. (Left) Bar chart showing cluster distribution sorted by descending size: Cluster 3 dominates (43.2\%, $\sim$259 images, likely interphase), while Cluster 1 comprises only 5\% ($\sim$30 images, likely apoptosis). (Right) PCA projection of feature space, coloured by cluster assignment, demonstrating substantial inter-cluster overlap.}
\label{figure:dist_unlabeled}
\end{figure*}

Prior to performing detailed analyses, we considered it essential to characterise the intrinsic structure and distribution of the full pool of available unlabelled data.
We performed an unsupervised clustering analysis on $N_U$ to better understand the intrinsic morphological distribution of the images. We evaluated multiple conventional feature extraction approaches and a blob detection feature extraction method \citep[e.g.][]{meijering12} provided robust cluster separation. This method segments the cells using Otsu thresholding \citep{otsu79}, identifies individual cell regions via connected-component and contour detection, and extracts shape-based statistics including object count, area, perimeter, and fill ratio. Cluster quality is then evaluated using the Silhouette score, which measures how similar each sample is to its assigned cluster relative to the nearest neighbouring cluster, with values ranging from 0 to 1 and higher scores indicating better-defined cluster separation. Using this metric, we obtained a Silhouette score of 0.382, suggesting moderate cluster structure in the data. The resulting cluster distribution (Figure~\ref{figure:dist_unlabeled}) exhibits class imbalance consistent with expected biological variability in cell-cycle phase durations. These features were projected into a lower-dimensional space using PCA and colour-coded for visualization, while $k$-means clustering was performed with $k=5$.

Although this unsupervised approach provides a global overview of the data organisation, substantial overlap between clusters is observed. This overlap likely reflects gradual biological transitions between cell-cycle stages, which complicate discrete separation and highlights the intrinsic difficulty of the classification task.

\subsubsection{Transfer Learning Data}

\textbf{CellCognition dataset:} To facilitate comparison with \citet{balluet22}, we used a dataset derived from the CellCognition software demonstration images \cite{held10}. The data consist of wide-field fluorescence time-lapse recordings of human HeLa Kyoto cells expressing histone H2B and $\alpha$-tubulin markers, visualising chromosomes and microtubules, respectively. Images were acquired at three distinct positions using a 20$\times$ dry objective with a temporal resolution of 4.6 minutes. We retained only the histone channel for analysis. The dataset comprises 1,011 images of size with most common dimension being $44 \times 51$ pixels, annotated across eight cell-cycle phases. The dataset spans eight cell-cycle phases with the following sample counts: Interphase (251), Prophase (112), Prometaphase (80), Metaphase (110), Early anaphase (40), Late anaphase (83), Telophase (136), and Apoptosis (199) 
(Fig.~\ref{fig:cell_cycle_samples_cog}).

\textbf{Homemade dataset:} From the perspective of embedding our algorithm in our smart microscope prototype, we prepared a dataset as detailed in \citet{bonnet24}. We used Hela Kyoto cells, whose DNA was labelled with Hoechst. Cells were imaged with a Zeiss inverted axio observer with a 20$\times$ dry objective. A double thymidine block synchronised the cells to get enough transient classes and equilibrate the dataset. We here selected contains 1321 cell images of size 72$\times$72 belonging to 7 different classes: anaphase (192), interphase (192 images), metaphase (192), prophase (192), prometaphase (192), telophase(192) and junk (169) (Figure \ref{figure:homemade}).

\textbf{RHC dataset}: We tested the generalisability by testing our network using published datasets in other contexts \citep{nagao20}. It features mouse retinal pigment epithelium cells (RPE1) with Hoechst (DNA) and CENP-F (centromeres) labellings. Cells were imaged with an Olympus IXplore SpinSR, equipped with a 60$\times$ oil-immersion objective. We summed up both channels before normalising. This dataset contains 461 original images of size 150$\times$150 belonging to 2 cell cycle phases: G2 (230 images, class 0) and non-G2 (231 images, class 1) Figure \ref{fig:RHC_HHG_HHE}. 

\textbf{HHG dataset}: It features human cervical cancer cells (HeLa) with Hoechst (DNA) and GM130 (Golgi) labellings. Cells were imaged with an Olympus IXplore SpinSR, equipped with a 60$\times$ oil-immersion objective. We summed up both channels before normalising. This dataset contains 491 original images of size 135$\times$135 belonging to 2 cell cycle phases: G2 (239 images, class 0) and non-G2 (252 images, class 1) Figure \ref{fig:RHC_HHG_HHE}. 

\textbf{HHE dataset}: It features human cervical cancer cells (HeLa) with Hoechst (DNA) and EB1 (microtubules plus-ends) labellings. Cells were imaged with an Olympus IXplore SpinSR, equipped with a 60× oil-immersion objective. We summed up both channels before normalising. This dataset contains 501 original images of size 128$\times$128 belonging to 2 cell cycle phases: G2 (258 images, class 0) and non-G2 (243 images, class 1) Figure \ref{fig:RHC_HHG_HHE}. 

\textbf{Ciliated cells dataset}: To further challenge our algorithm, we use an entirely distinct biological question, moreover using fixed cells. This dataset features mouse embryonic fibroblasts NIH3T3 with Hoechst (DNA) and acetylated-tubulin  (cilium) labellings. Cells were imaged with an Olympus IXplore SpinSR, equipped with a 60× oil-immersion objective. We summed up both channels before normalising. This dataset contains 558 original images of size 135$\times$135 belonging to 2 states: ciliated (279 images, class 0) and non-ciliated cells (279 images, class 1) (Figure \ref{fig:ciliated}).

\textbf{DIC dataset:} We further challenged the model on a non-fluorescence imaging modality: differential interference contrast (DIC). In DIC images, class-discriminative information is conveyed less by intensity than by optical path length gradients, producing edge-enhanced, relief-like patterns that emphasise texture and subtle morphological transitions rather than fluorescent signal localisation. We assembled a set of nematode one-cell embryos imaged during mitosis by using a Zeiss Axio-imager microscope. It features 169 original images of size 512$\times$512 belonging to 3 different classes: before nuclear envelope breakdown (NEBD, 57 images), metaphase (57), and anaphase (55) (Figure \ref{figure:dic}).

\section{Method}

We introduce SGAN, a semi-supervised deep learning framework for cell-cycle classification that exploits limited labelled data, abundant unlabelled data, and generated samples to improve decision-boundary optimisation. An overview of the proposed SGAN architecture is illustrated in Figure \ref{figure:sgan_arc}. 

We consider a classification task on cell microscopy images. Let $x \in \mathbb{R}^{H \times W \times C}$ denote a single microscopy image of size $H \times W$ pixels with $C$ channels (in our case, $C=1$), representing one of five cell cycle phases. The classification target is denoted as $y \in \{1, 2, ..., K\}$ where $K=5$ represents the five cell cycle phases: Anaphase, Interphase, Apoptosis, Metaphase, and Prometaphase from the Mitocheck data. During training of SGAN, labelled images provided ground-truth phase annotations, enabling the discriminator to learn class-conditional decision boundaries for the classification task, expressed as the posterior probability $P(y|x)$. For a given training dataset, we have:
\begin{enumerate}
	\item \textbf{labelled samples}: $\mathcal{L} = \{(x_i, y_i)\}_{i=1}^{N_L}$ where we evaluate four labelled data budgets: $N_L \in \{100, 200, 300, 400\}$ samples corresponding to $\{20, 40, 60, 80\}$ images per class across five cell cycle phases    
	\item \textbf{unlabelled samples}: $\mathcal{U} = \{x_j\}_{j=1}^{N_U}$ where $N_U = 600$ represents the total pool of unlabelled images.
    
   	 \item \textbf{Validation set}: $\mathcal{V} = \{(x_k, y_k)\}_{k=1}^{N_V}$ where $N_V = 0.2 \times N_L$ (20\% stratified split).
    
   	 \item \textbf{Test set}: $\mathcal{T} = \{(x_m, y_m)\}_{m=1}^{N_T}$ where $N_T = 100$ (fixed to 20 images per class) reserved for final evaluation.
\end{enumerate}

\subsection{Semi-Supervised GAN Architecture}

The SGAN couples supervised classification with unsupervised feature learning through a dual-head discriminator. A shared convolutional backbone extracts features from input images, which are then fed into (1) a supervised head performing multi-class cell-cycle phase classification using a softmax activation and sparse categorical cross-entropy loss, and (2) an unsupervised head that distinguishes real from generator-synthesized samples using a custom activation function (equation~\ref{equation:custom_act}, \citealt{salimans16}) and binary cross-entropy loss. This joint training encourages the emergence of a unified feature space that is both discriminative for classification and sensitive to the realism of generated data.

\begin{figure*}
\centering
\includegraphics[width=.95\linewidth]{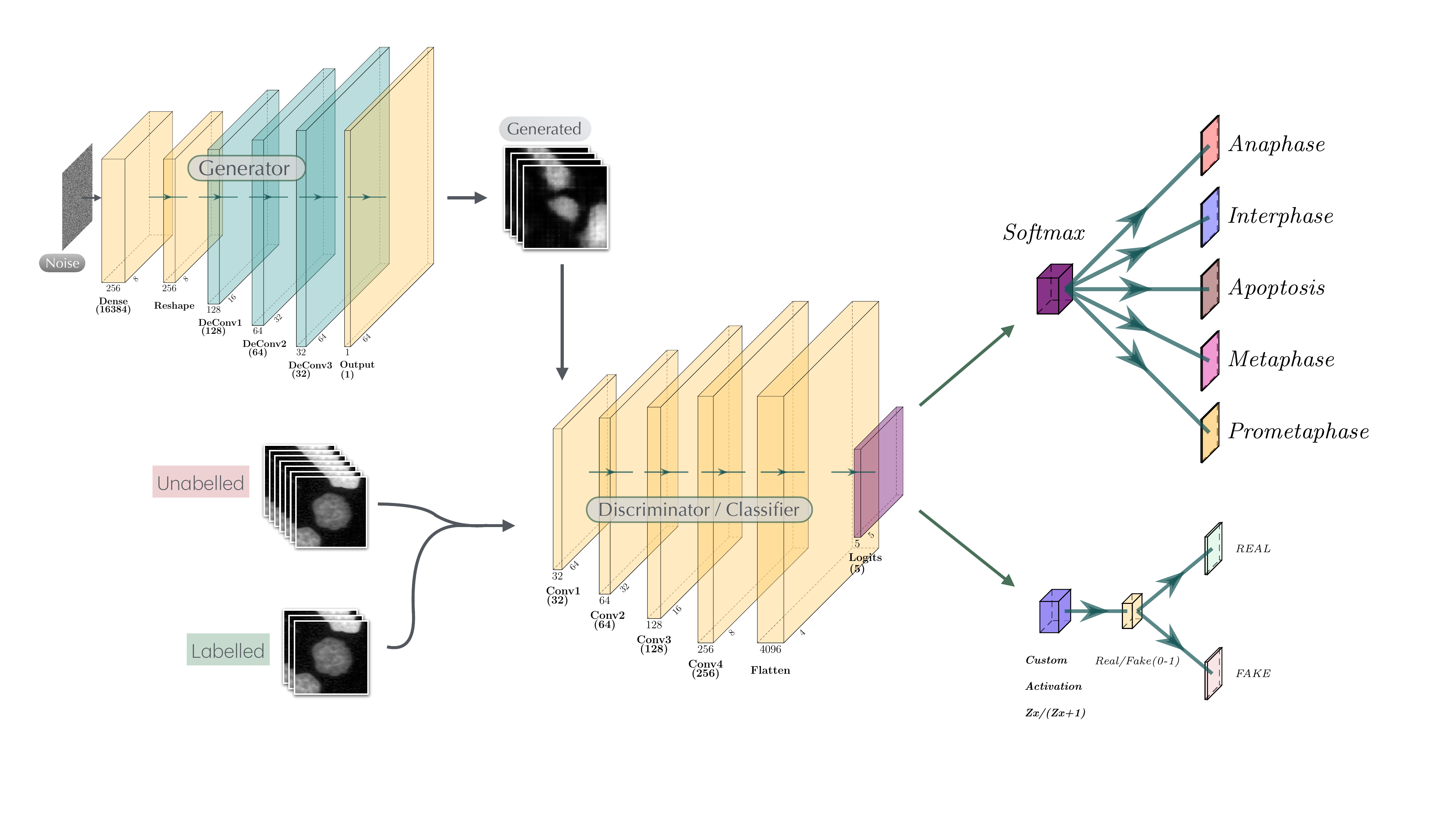} 
\caption{Architecture of the proposed Semi-Supervised GAN (SGAN). The generator maps random noise through dense and transposed convolutional layers to produce synthetic cell images. The discriminator/classifier receives real (labelled and unlabelled) and generated images, processes them through successive convolutional layers, and outputs (i) class probabilities via a softmax layer for multi-class cell-cycle classification and (ii) a real/fake prediction through a custom activation for adversarial training.}
\label{figure:sgan_arc}
\end{figure*}

Unlike conventional GANs, where the adversarial objective is limited to real-versus-fake discrimination \citep{goodfellow14b}, SGAN leverages this process to shape classification decision boundaries. As the generator progressively produces more realistic samples, it effectively explores the periphery of the true data distribution. In response, the discriminator, through its shared feature extractor, must learn increasingly refined representations to distinguish genuine data from synthetic samples. This interaction drives the model to identify low-density regions between class clusters, where class boundaries are most naturally defined. Consequently, the adversarial dynamics guide the feature space toward well-separated, generalisable decision boundaries, enabling SGAN to use generative competition not only for data synthesis but as a principled mechanism for semi-supervised classification. Figure~\ref{figure:sgan_arc} illustrates the complete architecture.

\subsubsection{Discriminator Architecture}

The discriminator consists of a shared feature extractor 
$D_{\text{base}}(\cdot; \theta_d)$ followed by two task-specific heads: 
(i) a supervised classifier trained on labelled data and 
(ii) an unsupervised real/fake discriminator trained on unlabelled and 
generated samples. The shared backbone comprises four convolutional blocks 
that progressively reduce spatial resolution while increasing channel depth, 
yielding a compact feature representation used by both heads. Joint 
optimisation of supervised and unsupervised objectives encourages the 
backbone to learn morphological features that are both discriminative for 
the five cell-cycle classes and effective for real/fake discrimination.

The supervised head maps shared features to class probabilities:
\begin{equation}
P(y \mid x; \theta_d) 
= D_{\text{sup}}(x) 
= \text{softmax}(D_{\text{base}}(x)) 
\in \mathbb{R}^{K},
\end{equation}
where $K$ denotes the number of classes.

It is trained on labelled images using sparse categorical cross-entropy loss:
\begin{equation}
\ell_{\text{sup}}
= -\sum_{i=1}^{N_L} \log P(y_i \mid x_i; \theta_d),
\end{equation}
where $\{(x_i, y_i)\}_{i=1}^{N_L}$ denotes the labelled training set.

The unsupervised head follows the formulation introduced by \citet{salimans16}, converting the shared logits into a scalar real/fake 
probability:
\begin{equation} \label{equation:custom_act}
D_{\text{unsup}}(x; \theta_d)
= \sigma_{\text{custom}}(D_{\text{base}}(x))
= \frac{Z_x}{Z_x + 1},
\end{equation}
with
\[
Z_x = \sum_{k=1}^{K} \exp\!\left(D_{\text{base}}(x)_k\right).
\]
This transformation maps the multi-class logits to the interval $(0,1)$ 
while preserving the structure of the shared feature representation.

The unsupervised discriminator is trained using binary cross-entropy loss on 
real unlabelled samples and generator-produced fake images:
\begin{equation}
\begin{aligned}
\ell_{\text{d, real}}
&= -\sum_{j=1}^{N_U} 
\log D_{\text{unsup}}(x_j^{\text{real}}; \theta_d), \\
\ell_{\text{d, fake}}
&= -\sum_{j=1}^{N_U} 
\log \!\left( 1 - D_{\text{unsup}}(x_j^{\text{fake}}; \theta_d) \right),
\end{aligned}
\end{equation}

The generator and discriminator are trained using batch size 
\[
N_B = \min\!\left(100, \left\lfloor \frac{N_{\text{labelled}}}{2} \right\rfloor \right),
\]
where $N_{\text{labelled}}$ is the total number of labelled training samples. The unsupervised discriminator head processes mini-batches of size $N_U = N_B/2$, alternating between real unlabelled and generated samples in sequential forward-backward passes. The supervised head uses batch size $N_S = N_B$ on labelled data. This sequential training approach ensures both real and generated distributions contribute to the shared feature extractor through consecutive gradient updates, while maintaining clean signal separation per data type.

\subsubsection{Generator Architecture}

The generator $G(\cdot; \theta_g)$ learns the feature distribution in the real images by transforming latent vectors into synthetic samples:

\begin{equation}
\tilde{x} = G(z; \theta_g) = \tanh\bigl(f_4 \circ f_3 \circ f_2 \circ f_1(z)\bigr),
\end{equation}

The generator consists of four transposed-convolution (stride-2) upsampling blocks 
$f_i$, which progressively increase the spatial resolution of the feature maps, 
followed by a final $\tanh$ activation constraining the output to $[-1,1]$. 
At each training iteration, a mini-batch of latent vectors 
$z \sim \mathcal{N}(0, I_d)$ with dimensionality $d=500$ 
is independently sampled and fed to the generator.
Resampling at every iteration ensures optimisation over the full latent support, 
promoting stable training and preventing overfitting to a fixed subset of latent inputs. 
Both the generator and discriminator include intermediate dropout layers 
($p=0.25$) to reduce overfitting and improve generalisation.

\subsection{Multi-Task Learning}

During each training iteration, the shared feature extractor $D_{\text{base}}$ is updated under three complementary objectives arising from the dual-head discriminator. Gradients from the supervised classification loss and from both unsupervised real and fake discrimination losses are applied successively to the same parameter set $\theta_d$. The effective update can be expressed as:

\begin{equation}
\nabla_{\theta_d} \mathcal{L}_{\text{total}}
=
\nabla_{\theta_d} \mathcal{L}_{\text{sup}}
+
\nabla_{\theta_d} \mathcal{L}_{\text{d,real}}
+
\nabla_{\theta_d} \mathcal{L}_{\text{d,fake}} .
\end{equation}

This multi-task formulation enforces that the shared features simultaneously optimise three objectives:
(1) discriminating between the five cell-cycle phases using limited labelled data,
(2) separating real unlabelled cell images from generator-produced samples, and
(3) progressively refining the classification decision boundaries in response to the evolving distribution of increasingly realistic synthetic images generated during adversarial training.

As a result, the shared feature space is shaped by both supervised labels and the broader structure present in the unlabelled data, enabling the model to capture morphological patterns that extend beyond the limited labelled dataset.

\section{Experimental Setup and Data Budget Analysis} \label{sec:ssgan}

We conducted a grid-search experiment to assess model accuracy across varying amounts of labelled and unlabelled data. The study evaluated all pairwise combinations of labelled data budgets (20, 40, 60, and 80 images per class) and unlabelled data pools (100, 185, 200, 300, 400, 500, and 600 images), yielding 28 distinct experimental configurations. We note that the 185-image unlabelled subset was stratified to maintain approximately equal class representation ($\sim$37 images per class).

Each configuration was trained using a standardised protocol. Models were trained for \(N_{\text{epochs}}\) ranging from 800 to 1500, with larger budgets assigned to lower-data regimes to compensate for increased training instability when fewer samples are available. The early-stopping patience was set adaptively between 10\% and 20\% of \(N_{\text{epochs}}\), depending on the data regime. This choice reflects the inherently unstable early dynamics of SGAN training, during which the generator and discriminator progressively reach equilibrium. Data augmentation was performed on the fly during training using rotations up to \(\pm 30^\circ\), translations up to 10\% of image width and height, brightness scaling in \([1.0,\,1.3]\), and zoom factors between 0.9 and 1.1. These settings were kept identical across all datasets and experiments, including the transfer-learning analyses in Section~\ref{sec:transfer_learning}.

Model checkpoints exceeding 75\% validation accuracy were automatically saved to enable post hoc selection of the best-performing model. The selected model was then evaluated on a held-out test set. To further assess robustness and estimate confidence intervals, performance was additionally examined using 5-fold stratified cross-validation on the labelled data.

All experiments were conducted on an internal high-performance computing server equipped with dual Intel Xeon Gold 6326 processors (64 cores, 128 threads total) and 256\,GB RAM. SGAN training used the Adam optimiser with learning rate \(\eta = 6 \times 10^{-4}\) and momentum parameter \(\beta = 0.5\), requiring a median training time of approximately 17\,min per configuration. 

\section{Results}

\subsection{SGAN Model Results}

We recently proposed the Roboscope, an autonomous microscope designed to capture rare and transient events \citep{bonnet24}. At its core is an image-analysis pipeline that detects objects of interest in real time. Developing such a smart microscopy system required addressing three key challenges: the scarcity of annotated training data, domain discrepancies between training and testing image distributions, and the need for lightweight models compatible with real-time decision-making on embedded hardware. Reported human expert accuracy in microscopy-based image classification is strongly task-dependent, ranging from approximately 50\% in challenging phenotype-recognition settings to around 70--75\% in more structured expert-driven classification tasks \citep[e.g.,][]{buetti-dinh19,Shpilman17}. On this basis, we set a target performance of at least 80\%, with the aim of matching or exceeding typical human expert performance in related applications.

Inference was computationally efficient. Single-cell patches of size 64$\times$64 pixels were classified in a mean time of 109.7\,ms, corresponding to a throughput of 8.9 cells per second. On the complete test set of 100 images, classification required 11.2\,seconds while achieving $93 \pm 2\%$ accuracy in the best case (Figure \ref{figure:accuracy_grid}). In a practical deployment scenario, processing a full 2048$\times$2048 pixel Roboscope field of view containing approximately 250 cells would require about 28\,seconds on standard CPU hardware, and this latency could be reduced even further with GPU acceleration.

\subsubsection{Impact of Labelled and Unlabelled Data}

\begin{figure}
\centering
\includegraphics[width=0.9\linewidth]{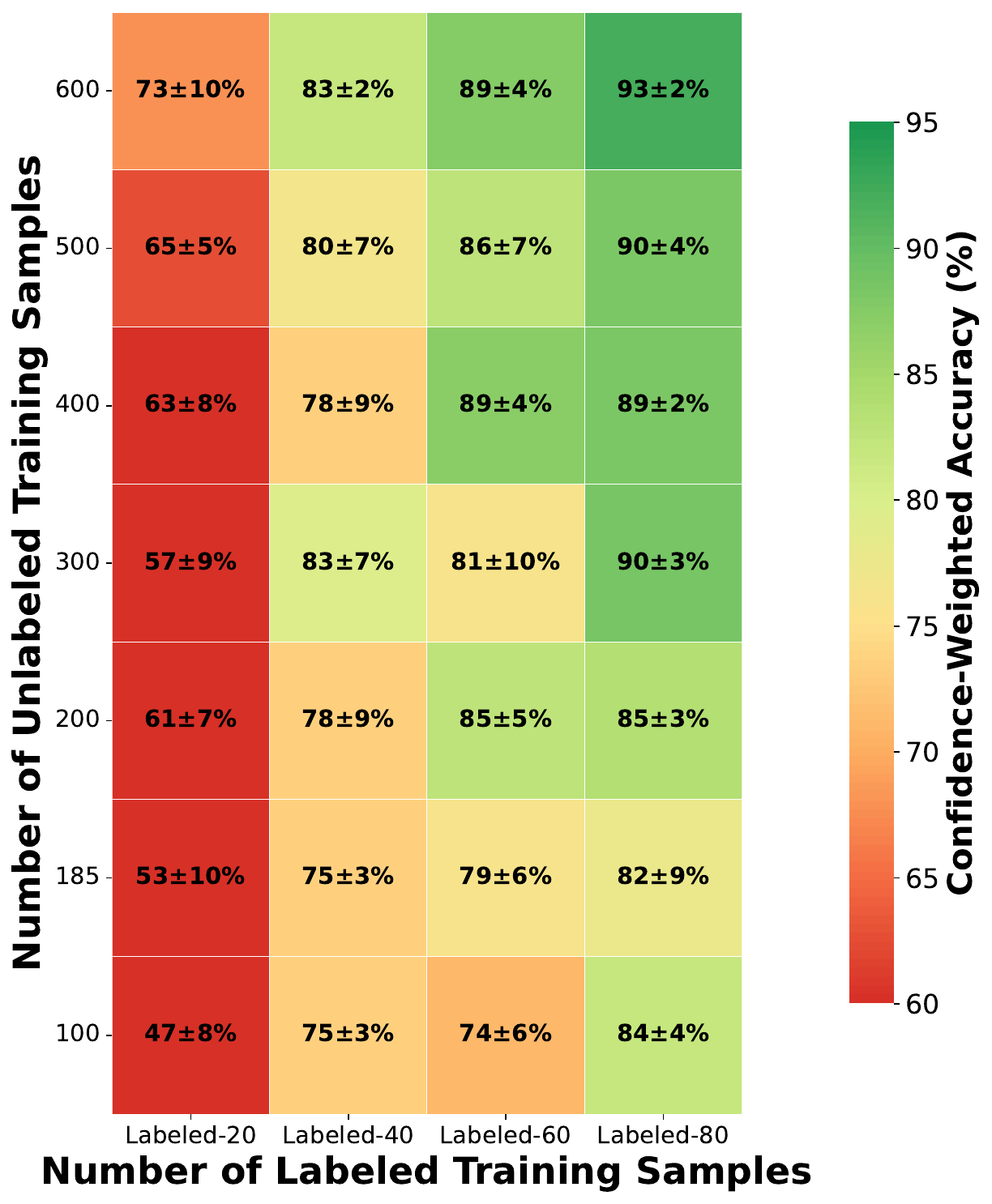} 
\caption{SGAN Grid Search: labelled-unlabelled data Trade-off Matrix showing mean test accuracy for all 28 configurations. Colors denote confidence-weighted accuracy. } 
\label{figure:accuracy_grid}
\end{figure}

To assess SGAN robustness under limited supervision, we systematically varied the numbers of labelled and unlabelled training samples. As shown in Figure~\ref{figure:accuracy_grid}, performance improved monotonically with both, but gains from labelled data were markedly larger. Increasing the number of labelled samples per class yielded substantial accuracy improvements ($\sim$20--40\%), whereas enlarging the unlabelled pool produced more modest gains ($\sim$5--10\%), except in the most extreme low-label regime. These results suggest that labelled data primarily define class-discriminative decision boundaries, while unlabelled data mainly regularise the representation by capturing the global data structure. Consistent with this interpretation, explicit class stratification of the unlabelled subset did not systematically improve performance.

Representative training curves are shown in Figure~\ref{fig:sgan_training_curves} for four configurations spanning low- and high-unlabelled-data regimes. These examples illustrate how sample availability affects convergence behaviour, training stability, and the balance between discriminator and generator dynamics during optimisation.

\subsubsection{Class Balance Effects In Unlabelled Data}

Comparison between the 185-image class-balanced subset and unstructured unlabelled pools of comparable size did not reveal a consistent performance benefit from explicit class stratification. For instance, with 80 labelled images per class, the stratified 185-image subset achieved an accuracy of $82 \pm 9\%$, which is slightly lower than that obtained with a 200-image unstructured pool ($85 \pm 3\%$). While this difference remains within the associated uncertainty, these results suggest that, in this regime, the SGAN learning objective is relatively robust to moderate class imbalance in the unlabelled data, provided that sufficient labelled samples are available to guide the supervised component.

\subsection{Comparison To Other Methods} \label{sec:comparison_methods}

\subsubsection{Transfer learning With Pretrained CNNs}

Deep CNNs have shown strong performance for cell-cycle phase classification in fluorescence microscopy, particularly with nuclear, Golgi, or microtubule staining \citep{acharya24,nagao20}. However, their performance generally depends on the availability of sufficiently large annotated datasets, which are often limited in bio-imaging. Transfer learning from models pre-trained on large-scale image datasets offers a common strategy to alleviate this constraint \citep{bayramoglu16,vonchamier21,yosinski14}.

To benchmark our SGAN framework, we implemented transfer learning by fine-tuning two ImageNet-pretrained architectures: MobileNetV2 \citep[3.37M parameters;][]{sandler18} and InceptionV3 \citep[27M parameters;][]{szegedy16}. These models were selected to represent different model sizes relevant to computational microscopy applications. Both were fine-tuned on our cell-cycle classification task using progressively smaller labelled subsets of 20, 40, 60, and 80 images per class (100, 200, 300, and 400 images in total) and evaluated on the same independent test set of 100 images used for SGAN.

\begin{figure}
\centering
\includegraphics[width=\linewidth]{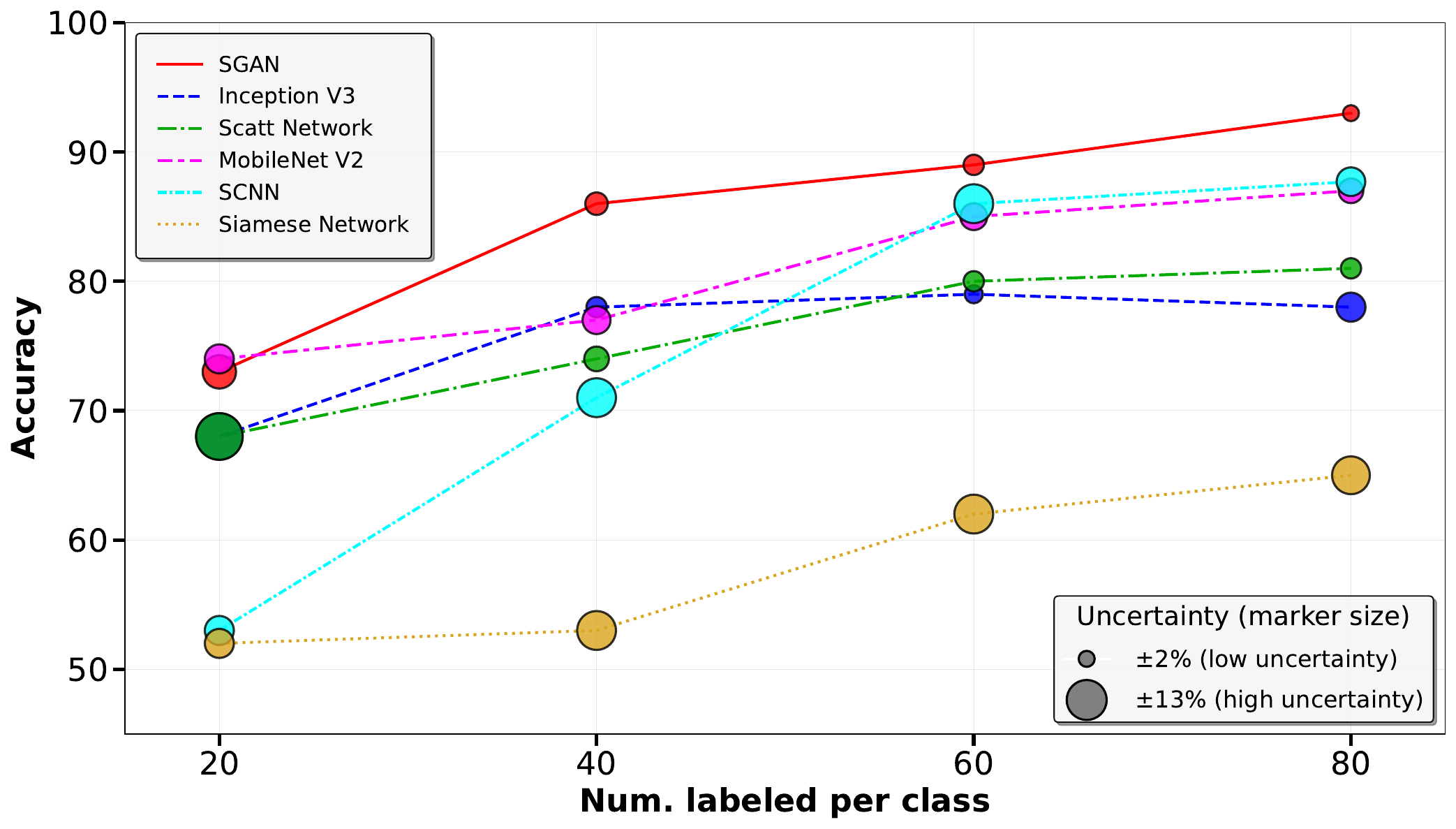} 
\caption{Performance comparison of different learning strategies on the Mitocheck dataset, including a few-shot learning Siamese network, a CNN trained from scratch, a scattering network, transfer learning with pre-trained models (InceptionV3 and MobileNetV2), and our proposed SGAN. The figure reports accuracy on the test set of 100 images as a function of the number of labelled training images}
\label{figure:comparison_models}
\end{figure}


Among these baselines, MobileNetV2 showed progressively improving performance with increasing supervision, reaching $87 \pm 6\%$ with 80 labelled images per class (Figure~\ref{figure:comparison_models}). In contrast, InceptionV3 showed higher variance and earlier performance saturation, peaking at $79 \pm 3\%$.

\subsubsection{Low-Depth CNN Trained From Scratch}
We next considered a smaller convolutional network to reduce the number of parameters and thus data greediness. Such a trivial solution also promised fast classification from the perspective of smart microscopy.  The low-depth CNN implementation employs a lightweight three-block convolutional architecture trained from scratch on the Mitocheck data set. The model architecture consists of three convolutional blocks, each containing $3 \times 3$ convolutions with ReLU activations, batch normalization, $2 \times 2$ max-pooling with stride-2, and dropout ($p=0.25$) for regularisation, progressively downsampling from input resolution $75 \times 75 \times 3$ to $4 \times 4 \times 128$ feature maps. The flattened features pass through two fully-connected layers (256 and 128 units, dropout $p=0.5$) before the softmax classification layer outputting five cell cycle phase probabilities. The model is trained with categorical cross-entropy loss and Adam optimizer ($\eta=0.0005$, $\beta_1=0.9$, $\beta_2=0.999$) for up to 1000 epochs. The training dataset is split into 80\% training and 20\% validation with stratified random sampling to maintain class distributions, while test samples from a completely separate folder (20 images per class) remain held-out for unbiased evaluation. Figure \ref{figure:comparison_models} displays the model performance compared to rest.

\subsubsection{Scattering Network}

Wavelet scattering networks, proposed by Mallat and colleagues \citep{bruna13}, extract features via cascaded wavelet decompositions without learned weights, addressing data scarcity by directly implementing invariances typically achieved through augmentation. Early applications in medical imaging (Rakotomamonjy et al., 2014) achieved $>80\%$ accuracy on 10-class lung cancer detection with $<100$ images per class. Our scattering baseline employs Kymatio (PyTorch) with $J=2$ scales and $L=8$ orientations, generating $\sim$1280-dimensional scattering coefficients per $64 \times 64$ image via Morlet wavelets with guaranteed Lipschitz stability. These invariant features feed into a shallow classifier (two fully-connected layers: 512 and 256 units with ReLU, batch normalization, dropout $p=0.3$) trained with categorical cross-entropy loss and Adam optimizer ($\eta=0.0005$, early stopping, patience=100 epochs).

The model achieves a 81\% performance in the best case for 80 labelled images per class, while only 68\% accuracy in the worst with 20 labelled images per class, where it outperforms the SCNN and Siamese model (Figure \ref{figure:comparison_models}).

\subsubsection{Siamese Network}

Siamese networks \citep{koch15} provide an alternative strategy for classification with limited labelled data. Our implementation used triplet inputs consisting of an anchor, a positive sample from the same class, and a negative sample from a different class, with shared weights across the three branches. Training was performed by optimising a triplet loss with margin \(m=1.0\):

\begin{equation}
\mathcal{L}_{\text{triplet}} = \max(d_{\text{pos}} - d_{\text{neg}} + 1.0, 0),
\end{equation}

where \(d_{\text{pos}} = \sum(a-p)^2\) and \(d_{\text{neg}} = \sum(a-n)^2\) denote the squared Euclidean distances between anchor-positive and anchor-negative embeddings, respectively.

The embedding network produced 150-dimensional representations through four fully connected layers with L2 regularization, dropout, and final L2 normalization. Models were trained with Adam (\(\eta = 10^{-4}\), \(\beta = 0.5\)) for up to 200 epochs. Training was performed on the different labelled subsets, each containing 20, 40, 60, and 80 images per class, using an 80/20 train-validation split, with triplets generated dynamically. Learned embeddings were evaluated on the test set using a \(k\)-nearest neighbors classifier (\(k=3\)).

Performance remained limited, increasing only from 52\% with 20 labelled images per class to 65\% with 80 images per class (Figure~\ref{figure:comparison_models}). This lower performance likely reflects the difficulty of learning sufficiently discriminative embeddings from small labelled datasets, together with the fact that, unlike SGAN, the Siamese framework does not exploit unlabelled data.

\subsubsection{Comparison With Other Studies} \label{sec:comparison_studies}

We compared SGAN with two recent semi-supervised GAN-based methods, SPARSE \citep{manni25} and Triple-GAN \citep{li17}, under identical experimental conditions. All models were trained with the same labelled subsets (20, 40, 60, and 80 images per class) and 600 unlabelled images. The labelled data were split into 80\% for training and 20\% for validation for early stopping, and final performance was evaluated on a fully held-out test set of 100 images (20 per class). Uncertainty was estimated using five-fold stratified cross-validation, with results reported as standard deviations and 95\% confidence intervals.

The three methods differ in their semi-supervised strategies. SPARSE uses pseudo-labelling with image-to-image translation, whereas Triple-GAN relies on a three-network architecture trained in separate stages. In contrast, SGAN performs implicit semi-supervised learning through a shared discriminator-based feature extractor jointly optimised with three losses, allowing the classifier to directly benefit from adversarially learned representations without explicit pseudo-labelling.

\begin{figure}
\centering
\includegraphics[width=\linewidth]{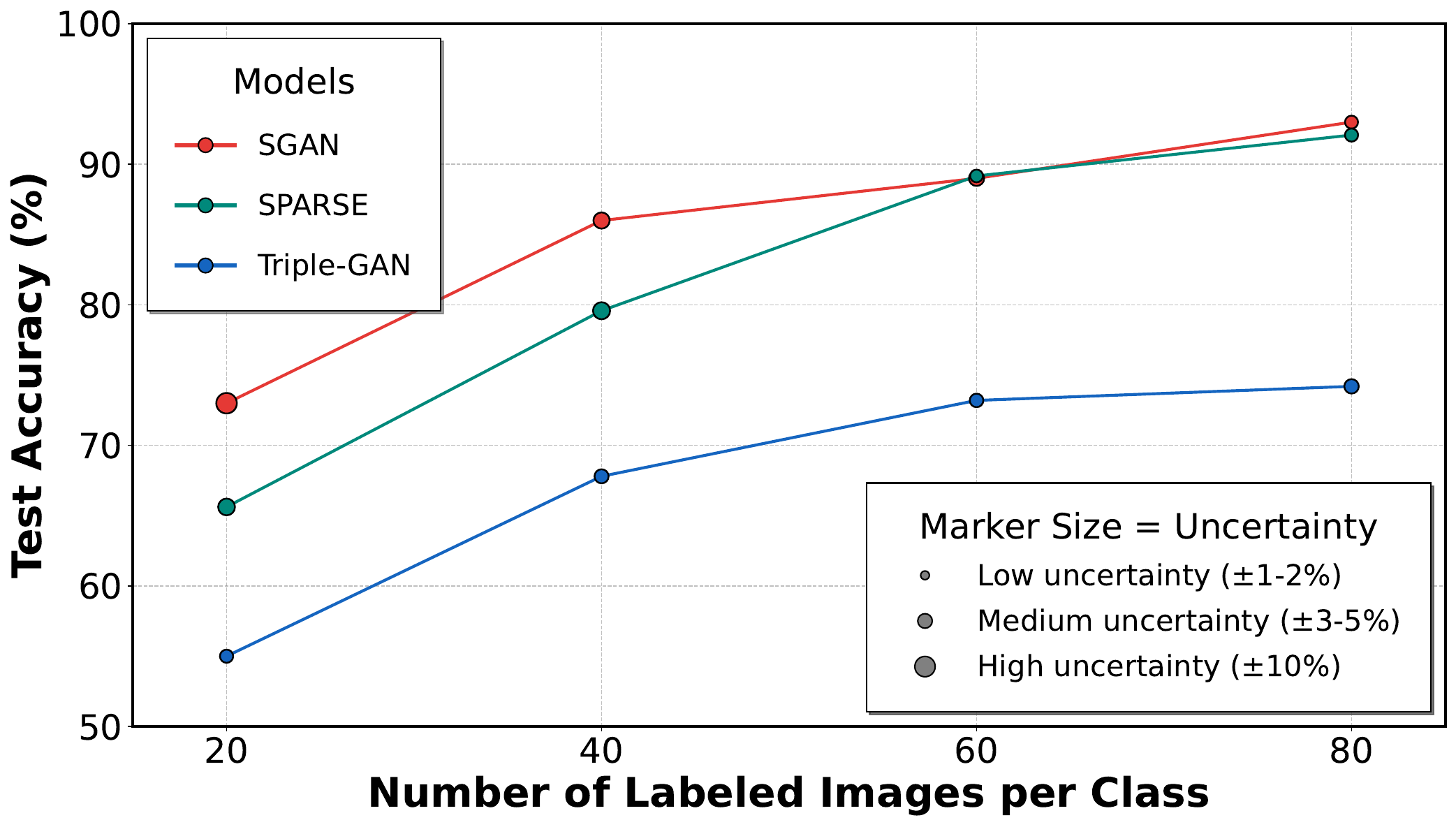} 
\caption{Comparative accuracy of semi-supervised cell phase classification models (SGAN, SPARSE, Triple-GAN) with varying amounts of labelled training data. All models used 600 unlabelled images. Marker sizes represent uncertainty from 5-fold cross-validation.}
\label{fig:comparison_lit}
\end{figure}

As shown in Figure~\ref{fig:comparison_lit}, SGAN generally provides the best or near-best performance across the evaluated data regimes, with the clearest advantage observed in the low-data setting where label efficiency is most important. SPARSE remains competitive and narrows the gap as more labelled data become available, whereas Triple-GAN tends to perform below both methods in the tested settings. Taken together, these findings suggest that SGAN is well suited to limited-supervision scenarios while retaining competitive performance at higher label budgets.


\subsection{Transfer Learning: Generalisability Of SGAN}

We also assessed the generalisability of SGAN across several cell classification tasks. Cell images are often visually similar, with only subtle differences between strains. Moreover, distinct imaged structures may still support classification \citep{nagao20}, since different proteins involved in the same process can equally reflect its progression. Some features may also be biologically redundant \citep{balluet22}. We therefore tested whether domain adaptation is feasible with SGAN despite its limited depth, using transfer learning, fine-tuning, and full retraining. To reflect realistic smart-microscopy applications, we considered shifts of increasing visual difference: strain, labelling, biological question (i.e. class meaning), and imaging modality (DIC). Concretely, we used the supervised component trained on Mitocheck to classify images from the CellCognition dataset, our homemade dataset, the datasets of \citet{nagao20}, and the DIC dataset.

\subsubsection{Data Preparation For Transfer Learning}

To improve robustness and generalisation across datasets, we applied the same augmentation pipeline used during SGAN training on Mitocheck data (Section \ref{sec:ssgan}). For most datasets (Cell\_cog\_5, Homemade Zeiss, HHE, HHG, RHC, and Cilia), this yielded an effective augmentation factor (= total images after augmentation/total number of original images) of \(3.8\times\) to \(4.0\times\).

Stronger augmentation was used for the 8-class CellCognition dataset because finer class granularity reduced the number of images per class and destabilised training; here, the augmentation factor was $\sim$4.5$\times$. The DIC dataset required the most aggressive augmentation because of its limited size (169 original images) and stronger morphological shift, with an augmentation factor of \(11\times\). Table~\ref{tab:dataset_image_counts} summarises the augmentation strategy across transfer-learning datasets.

\subsubsection{Transfer Learning On The CellCognition Dataset} \label{sec:transfer_learning}

We evaluated the transferability of the SGAN-pretrained five-class discriminator on the five-class CellCognition dataset (Cell\_cog\_5) \citep{held10}. Transfer learning was performed in two stages: first, a newly initialised classification head was trained with the convolutional backbone frozen; second, the last two to three convolutional layers were selectively fine-tuned using a reduced learning rate (\(\eta = 5 \times 10^{-5}\)). Class-weighted losses were used throughout to compensate for residual class imbalance.

Performance was assessed using stratified 5-fold cross-validation. In each fold, 20\% of the data were held out for testing, while the remaining 80\% were split into training (64\% of total) and validation (16\%) subsets. A fresh classification head was initialised for each fold to avoid weight leakage. In the five-class setting, this yielded a test accuracy of \(95.68 \pm 1.33\%\) (Table~\ref{tab:cellcog_5class_vs_8class_cv}). Per-class results were consistently high, with F1-scores above 0.95 for Interphase, Metaphase, and Apoptosis, and somewhat lower but still strong performance for Prometaphase and Anaphase. The low standard deviations across classes indicate stable transfer performance. Confusion matrices are shown in Figure \ref{fig:conf_transfer_5_8_class}. 

\begin{table*}[t]
\centering
\renewcommand{\arraystretch}{1.1}
\small
\begin{tabular}{lccc|cccc}
\hline
\multicolumn{4}{c|}{\textbf{5-Class CellCognition}} & \multicolumn{4}{c}{\textbf{8-Class CellCognition}} \\
\hline
\textbf{Class} & \textbf{F1-Score} & \textbf{Precision} & \textbf{Recall} & \textbf{Class} & \textbf{F1-Score} & \textbf{Precision} & \textbf{Recall} \\
\hline
Anaphase & $0.922$ & $0.938$ & $0.912$ & Early Ana. & $0.958$ & $0.960$ & $0.958$ \\
Apoptosis & $0.995$ & $0.991$ & $1.000$ & Late Ana. & $0.839$ & $0.826$ & $0.858$ \\
Interphase & $0.968$ & $0.953$ & $0.986$ & Class I & $0.894$ & $0.883$ & $0.907$ \\
Metaphase & $0.957$ & $0.972$ & $0.944$ & Junk (J) & $0.976$ & $1.000$ & $0.953$ \\
Prometaphase & $0.928$ & $0.928$ & $0.928$ & Metaphase & $0.805$ & $0.831$ & $0.783$ \\
\hline
 &  &  &  & Prometaphase & $0.959$ & $0.995$ & $0.927$ \\
 &  &  &  & Prophase & $0.863$ & $0.861$ & $0.867$ \\
 &  &  &  & Telophase & $0.866$ & $0.838$ & $0.898$ \\
\hline
\multicolumn{4}{c|}{Test Acc: $95.68 \pm 1.33\%$} & \multicolumn{4}{c}{Test Acc: $89.42 \pm 2.82\%$} \\
\hline
\end{tabular}
\caption{Per-class performance metrics comparison between 5-class and 8-class Cell Cognition datasets across 5-fold stratified cross-validation. The 5-class dataset achieved superior performance with $95.68 \pm 1.33\%$ test accuracy, while the 8-class dataset demonstrated robust transfer learning to expanded cell cycle phases with $89.42 \pm 2.82\%$ test accuracy. Error bars shown only for test accuracy; per-class metrics report mean values across folds.}
\label{tab:cellcog_5class_vs_8class_cv}
\end{table*}

To assess fine-grained transferability, we extended the task to eight classes (Cell\_cog\_8) by subdividing Anaphase into early and late stages and adding Prophase and Telophase. Despite the increased complexity, the model achieved strong performance with \(92.73\%\) test accuracy. However, it struggled more with Prophase and Telophase (Figure~\ref{fig:conf_transfer_5_8_class}), likely because these phases were absent from the original SGAN training data. Transfer learning remained lightweight, requiring only $\sim$2 min per fold for the five-class CellCognition dataset and $\sim$5 min for the eight-class variant. Overall, the modest drop from five to eight classes suggests that SGAN-learned features are generalisable, while highlighting the challenges of more fine-grained cell-cycle classification.

\subsubsection{Classification On Homemade Dataset}
We also evaluated transfer learning on 1321 homemade Zeiss microscopy images spanning seven cell-cycle phases. For the classification architecture, we leveraged the frozen pre-trained SGAN discriminator as a feature extractor, appended with a trainable classification head consisting of three dense layers (512, 256, and 128 neurons with ReLU activation and batch normalisation) followed by a softmax output layer for seven classes. The model was trained using the Adam optimiser ($\eta = 0.0003$, $\beta_1=0.9$, $\beta_2=0.999$) with class-balanced weights. These hyperparameters were selected to provide adaptive per-parameter learning rates while maintaining stable convergence on the limited transfer learning dataset.

\begin{table}[h]
\centering
\small{
\begin{tabular}{lccc}
\toprule
\textbf{Class} & \textbf{F1-Score} & \textbf{Precision} & \textbf{Recall} \\
\midrule
Anaphase (A) & $0.981$ & $0.992$ & $0.970$ \\
Interphase (I) & $0.810$ & $0.816$ & $0.805$ \\
Junk (J) & $0.964$ & $0.988$ & $0.941$ \\
Metaphase (M) & $0.960$ & $0.953$ & $0.969$ \\
Prophase (P) & $0.923$ & $0.901$ & $0.947$ \\
Prometaphase (PM) & $0.958$ & $0.956$ & $0.960$ \\
Telophase (T) & $0.878$ & $0.877$ & $0.880$ \\
\hline
\multicolumn{4}{c}{Test Accuracy: $92.41 \pm 1.4 \%$} \\
\hline
\end{tabular}
}
\caption{Per-class performance metrics on the 7-class Zeiss homemade microscopy dataset across 5-fold stratified cross-validation. }
\label{tab:7class_zeiss_results}
\end{table}

Following 5-fold stratified cross-validation, transfer learning achieved a mean test accuracy of \textbf{$92.41 \pm 1.40\%$} (Table~\ref{tab:7class_zeiss_results}), with stable performance across folds ($89.69\%$--$93.56\%$) and convergence in approximately $5$ minutes per fold. The model generalised well across most cell-cycle phases, with stronger performance for Anaphase, Metaphase and Prometaphase than for phases absent from the original SGAN training data, such as Prophase and Telophase (Figure~\ref{fig:conf_transfer_5_8_class}). This suggests that transfer learning from an SGAN pre-trained on incomplete cell-cycle data struggles with unseen phases, and that fine-tuning on a complete dataset or using architectures trained on all seven phases could further improve performance.

\subsubsection{Classification Under Alternative Cell-Cycle Labels}
\textbf{Nagao Binary Classification:} We also evaluated transfer learning on previously published microscopy datasets with alternative binary labelling schemes~\citep{nagao20}. Unlike Mitocheck, which primarily annotates mitotic stages, three datasets (HHG, RHC, and HHE) focus on G2 versus non-G2 classification, corresponding to a subphase-specific distinction within interphase. The Cilia dataset is biologically distinct, as it targets cilia-related status rather than cell-cycle phase, and was therefore considered as a complementary transfer task.

The four binary datasets (HHG, Cilia, RHC, and HHE) were constructed by combining two fluorescence channels per sample. After augmentation, each dataset contained 1,857--2,234 images (Table~\ref{tab:dataset_image_counts}) and was split into \(\sim\)60/20/20 train/validation/test sets with near-balanced classes (48.7--51.3\%). Transfer-learning results are reported in Table~\ref{tab:per_class_metrics}.

\begin{table}[h]
\centering
\resizebox{0.50\textwidth}{!}{
\begin{tabular}{ll|cc|cc}
\toprule
\textbf{Dataset} & \textbf{Class} & \textbf{F1} & \textbf{Prec.} & \textbf{Recall} & \textbf{Test Acc.} \\
\midrule
\multirow{2}{*}{HHG} & Class 0 & $0.896$ & $0.870$ & $0.926$ & \multirow{2}{*}{$89.60 \pm 0.75\%$} \\
& Class 1 & $0.895$ & $0.926$ & $0.868$ & \\
\midrule
\multirow{2}{*}{Cilia} & Class 0 & $0.884$ & $0.887$ & $0.882$ & \multirow{2}{*}{$88.45 \pm 0.71\%$} \\
& Class 1 & $0.885$ & $0.883$ & $0.887$ & \\
\midrule
\multirow{2}{*}{RHC} & Class 0 & $0.860$ & $0.895$ & $0.829$ & \multirow{2}{*}{$86.65 \pm 0.86\%$} \\
& Class 1 & $0.872$ & $0.843$ & $0.904$ & \\
\midrule
\multirow{2}{*}{HHE} & Class 0 & $0.802$ & $0.868$ & $0.748$ & \multirow{2}{*}{$81.09 \pm 4.91\%$} \\
& Class 1 & $0.819$ & $0.769$ & $0.878$ & \\
\midrule
\multirow{3}{*}{DIC} 
& Aan  & $0.873$ & $0.926$ & $0.826$ & \multirow{3}{*}{$84.13 \pm 7.42\%$} \\
& M    & $0.870$ & $0.812$ & $0.936$ & \\
& NEBD & $0.967$ & $0.992$ & $0.944$ & \\
\bottomrule
\end{tabular}
}
\caption{Transfer learning performance (5-fold stratified cross-validation). DIC results correspond to fine-tuned transfer learning on the three-class DIC dataset.}
\label{tab:per_class_metrics}
\end{table}

Performance depended on marker morphology and localisation stability. Among these binary transfer-learning tasks, HHE was the most challenging, reaching \(81.09 \pm 4.91\%\) test accuracy and showing the highest variability across folds, likely reflecting weaker structural detail and less distinctive morphological patterns (see Figure~\ref{fig:RHC_HHG_HHE}). By contrast, HHG achieved \(89.60 \pm 0.75\%\), consistent with the stable localisation pattern of the Golgi marker, while RHC reached \(86.02\%\) balanced accuracy with moderate variability. The Cilia dataset achieved \(88.45 \pm 0.71\%\), indicating that the transferred features remain effective even in a related but non-cell-cycle binary classification setting. Across the four datasets, transfer learning required on average only 3~min~30~s per fold. Overall, performance was largely driven by the distinctness and stability of marker-specific morphological features.

Comparison with \citet{nagao20} provides useful context. Nagao et al.\ used supervised CNNs to classify interphase cell-cycle states from fluorescence microscopy, reporting ~90\% accuracy for G1/S versus G2 discrimination with informative channels (Hoechst, nuclear area, DNA intensity, Golgi organisation). In contrast, our framework was trained in a semi-supervised, low-label setting via SGAN and adapted through transfer learning. Despite this more challenging setting, our transferred representation remained highly competitive across all four datasets from \citet{nagao20}, demonstrating the generalisability of the learned feature space beyond the original Mitocheck task.

\textbf{Classification on the DIC dataset:} To further assess generalisation, we evaluated the model on a markedly different imaging modality, differential interference contrast (DIC) microscopy. This setting introduces a pronounced domain shift: differences in resolution, imaging physics, contrast formation, and morphological cues challenge the feature representations learned from fluorescence-based Mitocheck data. We employed a custom \textit{C.~elegans} embryo dataset comprising three classes (169 original images in total). Each image was augmented with ten variants and the dataset was split into stratified training, validation, and test sets following a 60/20/20 protocol. 

Direct transfer learning from the Mitocheck pre-trained model yielded limited performance. By contrast, full retraining initialised from the Mitocheck weights substantially improved performance (Table~\ref{tab:per_class_metrics}). This improvement was achieved at a substantially greater computational cost, with retraining requiring approximately 1 h per fold (Figure~\ref{figure:conf_mat_dic}).

\section{Discussion}

We previously developed a cell-classification approach based on hand-crafted features and a machine-learning classifier \citep{balluet22}, fast enough for on-the-fly use in automated microscopy, but requiring full retraining and feature re-selection when imaging conditions changed. 

Here, we addressed these limitations using a semi-supervised GAN framework that combines automatic feature extraction with strong data efficiency under limited supervision. SGAN achieved competitive accuracies of \(90\text{--}93\%\) using only tens of labelled samples per class, substantially fewer than typically required by fully supervised approaches \citep{haque21, manni25, zhong25}. Varying the amounts of labelled and unlabelled data showed that labelled samples are the main driver of classification performance, while unlabelled data provide a complementary regularising effect, particularly in the low-label regime. The narrow variance bands in the cross-validation errors indicate stable and consistent behaviour across runs.

Benchmarking confirmed the strength of the approach. Under the same experimental setting, SGAN outperformed alternative methods, including pre-trained ImageNet-based models, InceptionV3, Siamese, scattering-based, and standard CNN baselines. It also compared favourably with recent semi-supervised GAN-based classifiers from the literature under matched experimental settings.

Transfer-learning experiments showed that the SGAN discriminator learns representations that generalise beyond the original Mitocheck domain. Strong performance on CellCognition and the 7-class Zeiss dataset indicates that these features capture morphological primitives that remain informative across microscopy set-ups and labelling schemes. However, the model showed slightly reduced performance with newly introduced classes such as Prophase and Telophase, which were absent from the original SGAN training framework. In most fluorescence-based tasks, fine-tuning only the final convolutional layers was sufficient. This adaptation remained lightweight, requiring on average only $\sim$4 min\,40\,s across datasets.

By contrast, the DIC experiments revealed the limits of this transferability, since the fundamentally different image-formation physics reduced the effectiveness of direct transfer and required broader retraining. The Nagao binary datasets provided an additional test of generalisability across cell type, marker, and biological question. Accuracies above \(86\%\) in most cases indicate that the SGAN backbone captures sufficiently rich cell-cycle related morphology even when the target task differs from the source labels. The lower and more variable performance on HHE suggests that transfer also depends on the visual distinctness and stability of marker-specific patterns.

A key strength is achieving these results with a relatively compact architecture. Unlike large pre-trained backbones, the SGAN discriminator remains lightweight while retaining strong transferability across domains. This combination of high accuracy (93\%), compact model size, and fast classification inference (109.7 ms per cell) makes the method particularly well-suited for real-time automated cell cycle classification on resource-constrained platforms such as the Roboscope.

\section{Conclusion}

We present a semi-supervised generative adversarial framework for cell-cycle classification under limited annotation constraints. SGAN enables accurate and data-efficient learning from microscopy images, while providing a lightweight foundation for real-time, event-driven autonomous microscopy.

Beyond its performance on the source task, the method showed good transferability across multiple fluorescence datasets, although the DIC experiments highlighted the remaining challenge of stronger modality shifts in imaging physics. 

Future work could further improve cross-domain robustness through domain-invariant representation learning, for example by using domain-adversarial objectives to reduce sensitivity to modality-specific cues. Better alignment between the generative and classification objectives, including adaptive or reward-based training schemes \citep[e.g.][]{siyuanli24}, may also improve stability and discriminative performance. In addition, large-scale self-supervised pre-training on unlabelled microscopy images could further enhance generalisation across cell types and imaging modalities while preserving data efficiency of the proposed method.

\section*{Data Availability}

The codes related to creating the SGAN model and transfer learning is available on Zenodo (DOI: 10.5281/zenodo.20445522).

The Mitocheck dataset that was used to train the SGAN model, is available in ~\citet{neumann10}.

For transfer learning validation, we evaluated SGAN generalisation on seven independent, datasets:

CellCognition dataset: 1,011 images available online in relation to publication~\citet{held10}.

Homemade dataset: 1,321 images available with preprint~\citep{bonnet24} and deposited in Zenodo (doi: 10.5281/zenodo.13832258).

Nagao datasets: Four datasets (RHC, HHG, HHE, and Ciliated cells) containing 461, 491, 501, and 558 images respectively, available from publication~\cite{nagao20} and deposited in Zenodo (doi: 10.5281/zenodo.3745864).

DIC dataset: 169 images from differential interference contrast (DIC) microscopy, first presented in this work and made publicly available through Zenodo (doi: 10.5281/zenodo.20024669).

\section*{Acknowledgements}

We acknowledge Prof. Thomas Walter for providing access to the Mitocheck dataset. 
This project was supported by Région Bretagne 
(AAP PME 2018--2019 -- Roboscope; iDémo 2024 -- Roboscope) 
and the Agence Nationale de la Recherche 
(PRCE project SAMIC, ANR-19-CE45-0011).

Some microscopy imaging was performed at the Microscopy Rennes Imaging Center 
(UMS 3480 CNRS/US 18 INSERM/University of Rennes). 
We acknowledge the France-BioImaging infrastructure supported by the 
French National Research Agency (ANR-10-INBS-04).

We also thank all members of the CeDRE, MFC teams, and 
Dr.~Remy Torro for valuable discussions.

\section*{CRediT authorship contribution statement}

Conceptualization: RM, YEH, JP, MT. 
Methodology: RM, YEH, MG, JP, CM, OB. 
Software: RM, YEH, CM, JP. 
Validation: RM, MG, CM, JP, MT. 
Formal analysis: RM, YEH, CM, JP. 
Investigation: RM, YEH, MG, CM, JB, LR, SP, JP. 
Resources: JB, LR, SP. 
Data curation: RM, JB, LR, JP, MT. 
Writing – Original Draft: YEH, JP. 
Writing – Review \& Editing: RM, JP, MT, CM. 
Visualization: RM, YEH, JP. 
Supervision: JP, MT. 
Project administration: JP, MT, OB. 
Funding acquisition: JP, MT, OB, OC.

\section*{Declaration of Competing Interest}

CM is employed by Inscoper, SAS, OC is the Chief Executive Officer and OB serves as the company's Chief Technical Officer. 
The Roul et al.\ (2014) patent on optimal driving is exclusively licensed to Inscoper, SAS. 
The company also co-owns, with CNRS and the University of Rennes, 
the Balluet et al.\ (2020) patent cited in this paper. 
JP and MT serve as scientific advisors to the company. 

The funders had no role in the study design, data collection and analysis, 
decision to publish, or preparation of the manuscript.
\bibliographystyle{cas-model2-names}
\bibliography{refs_new_clean.bib}

\appendix

\section{SGAN training curves examples}

\begin{figure}[!htbp]
    \centering
    
    \includegraphics[width=0.4\textwidth]{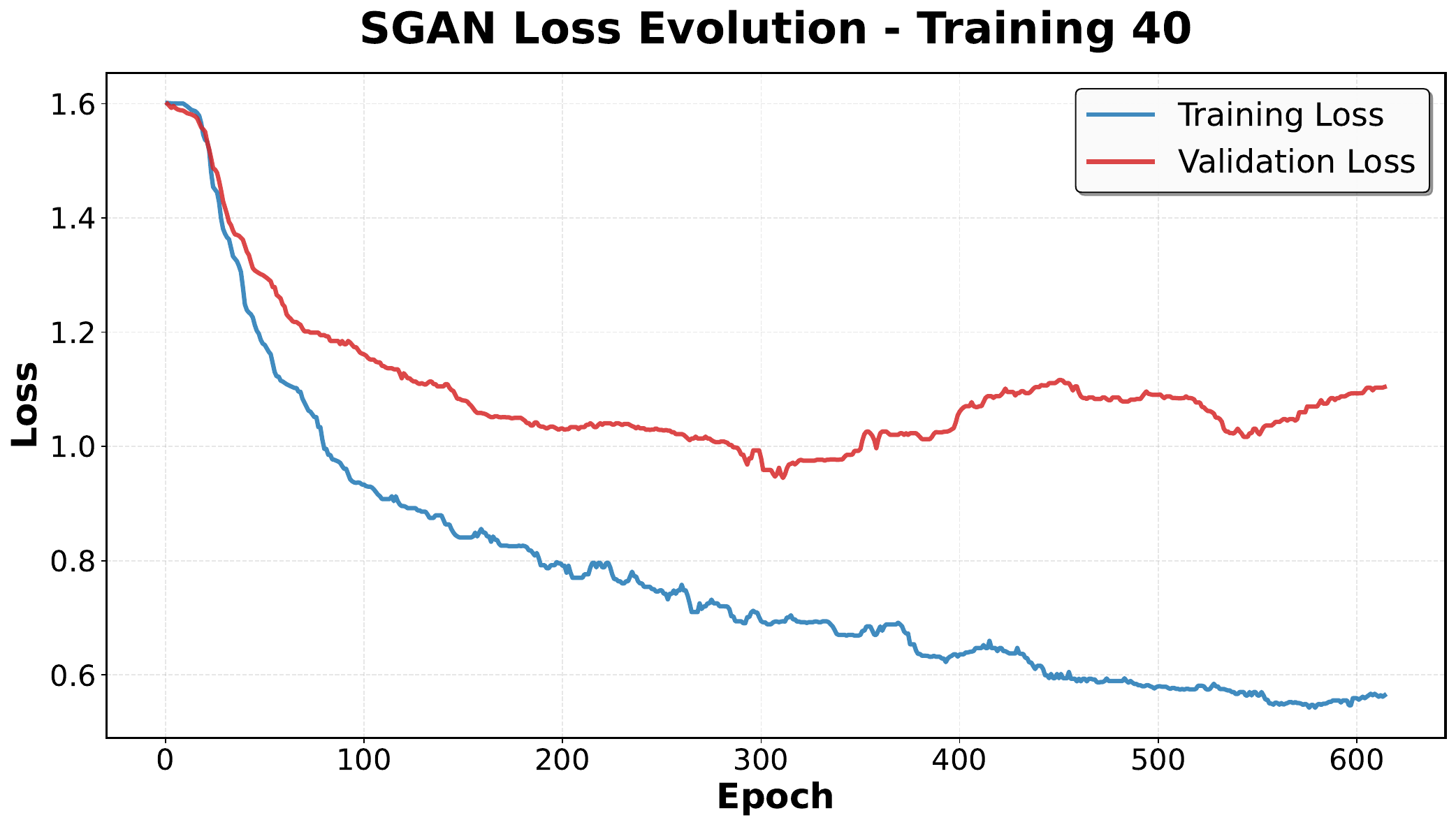}
    \includegraphics[width=0.4\textwidth]{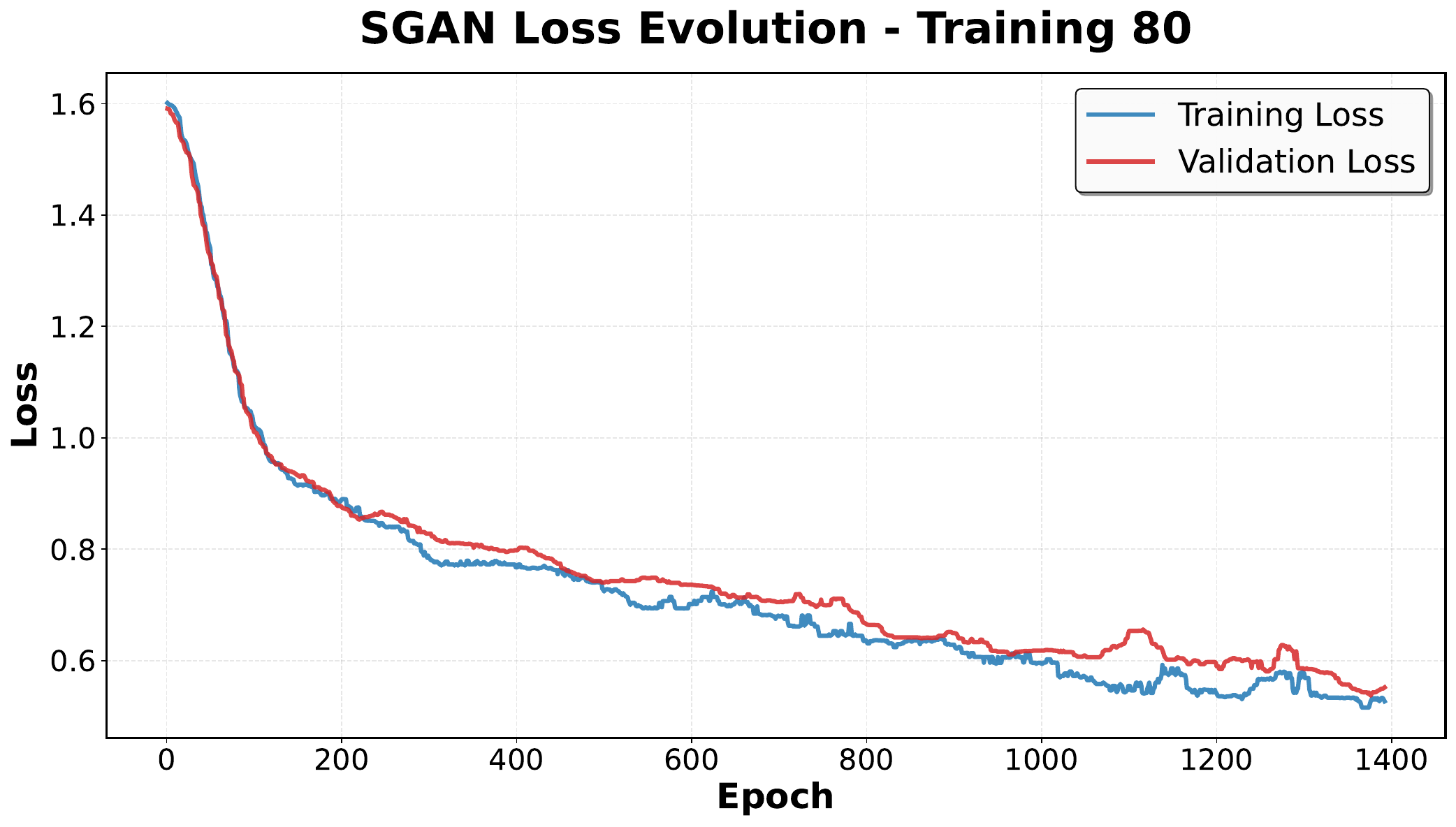}
    
%
    \includegraphics[width=0.4\textwidth]{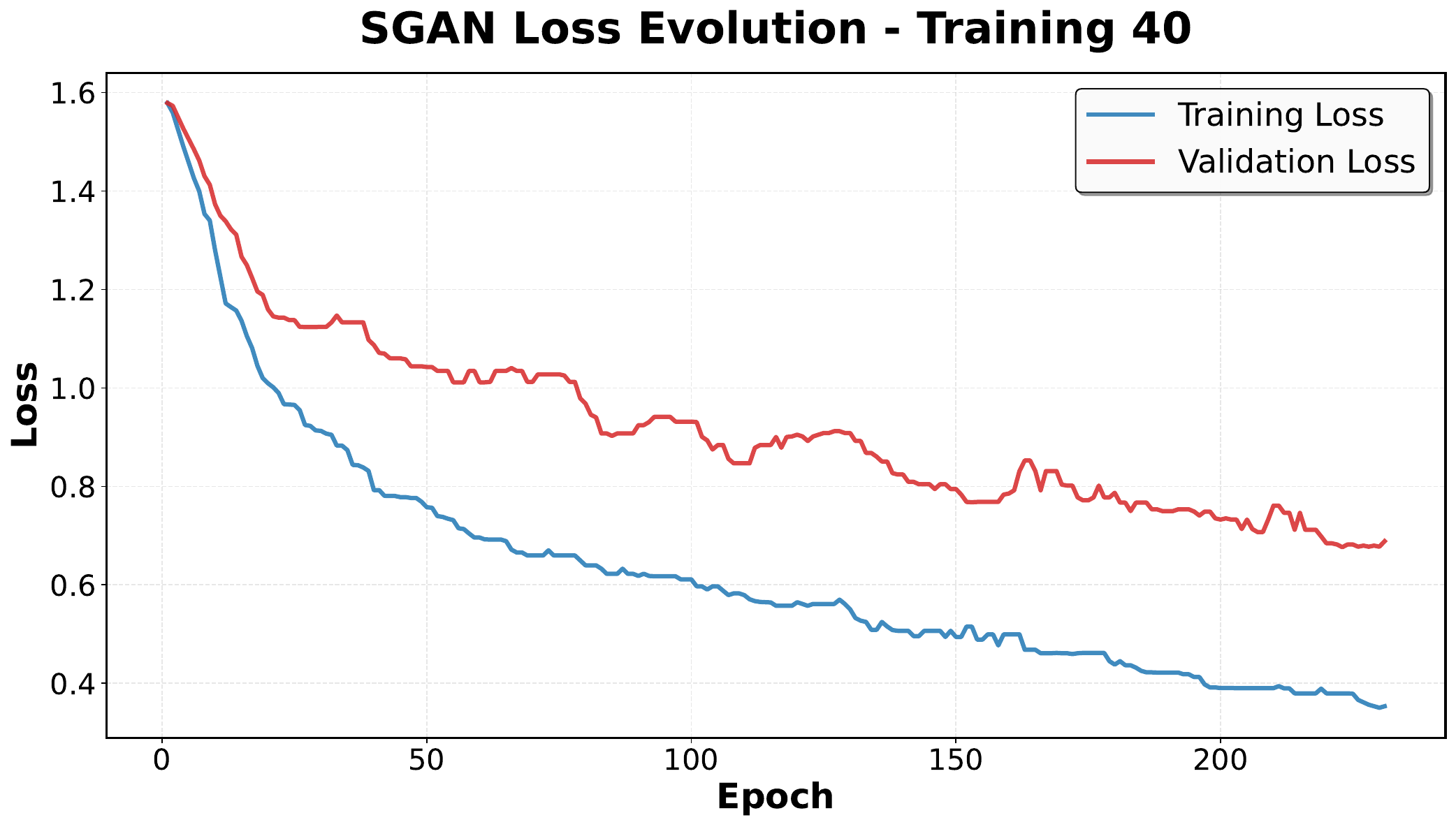}
    \includegraphics[width=0.4\textwidth]{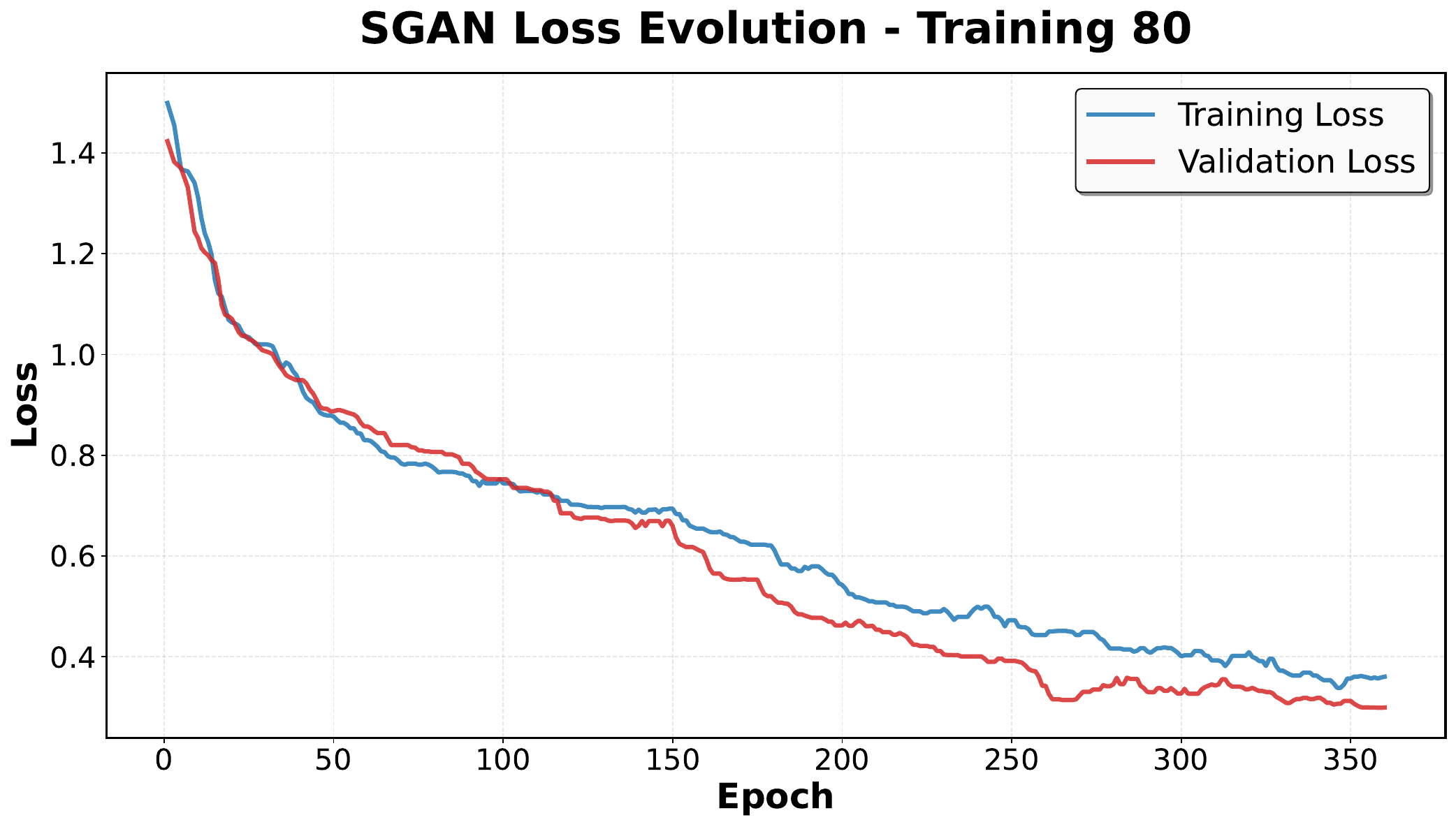}
    
    \caption{SGAN classifier training loss evolution under different labelled and unlabelled data regimes. Top row: $U=100$ with 40 and 80 labelled samples. Bottom row: $U=600$ with 40 and 80 labelled samples. Overfitting is observed in regimes with limited labelled data.}
    \label{fig:sgan_training_curves}
\end{figure}
\clearpage

\section{Confusion matrices of Transfer Learning results on CellCognition and Homemade Zeiss data sets.}

\begin{figure}[!h] 
\centering
\includegraphics[width=.9\linewidth]{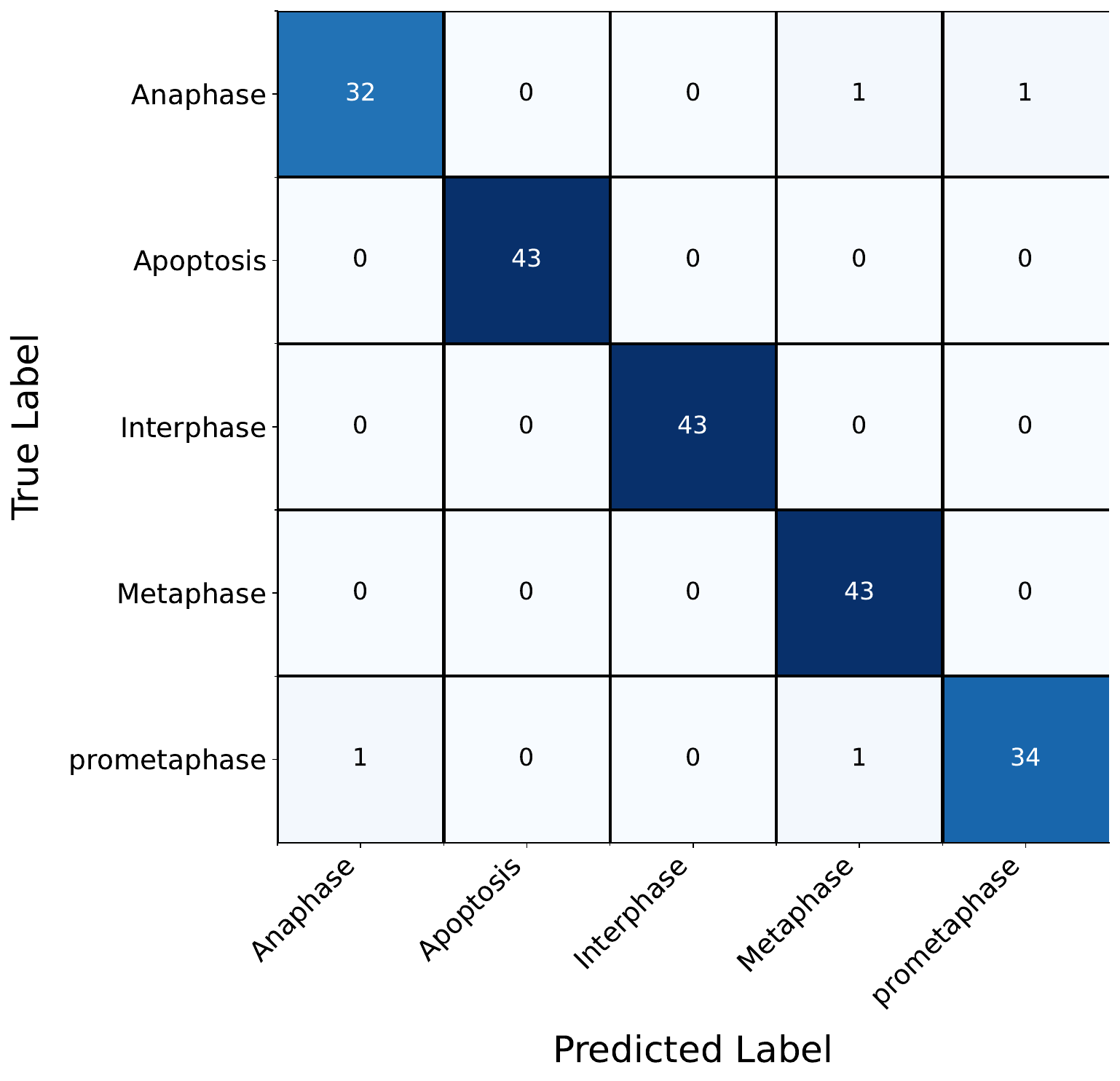} 
\includegraphics[width=.9\linewidth]{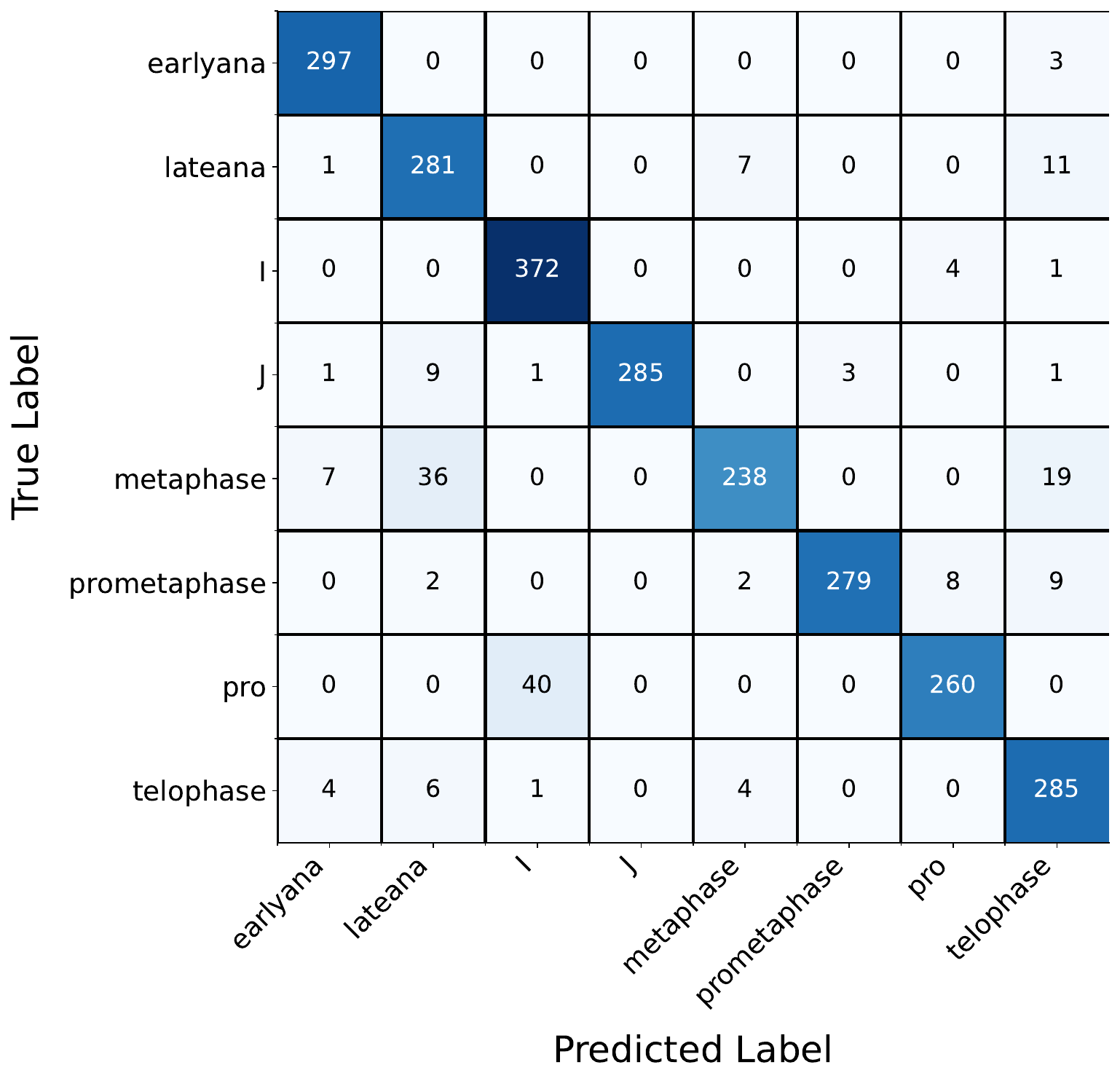} 
\includegraphics[width=.9\linewidth]{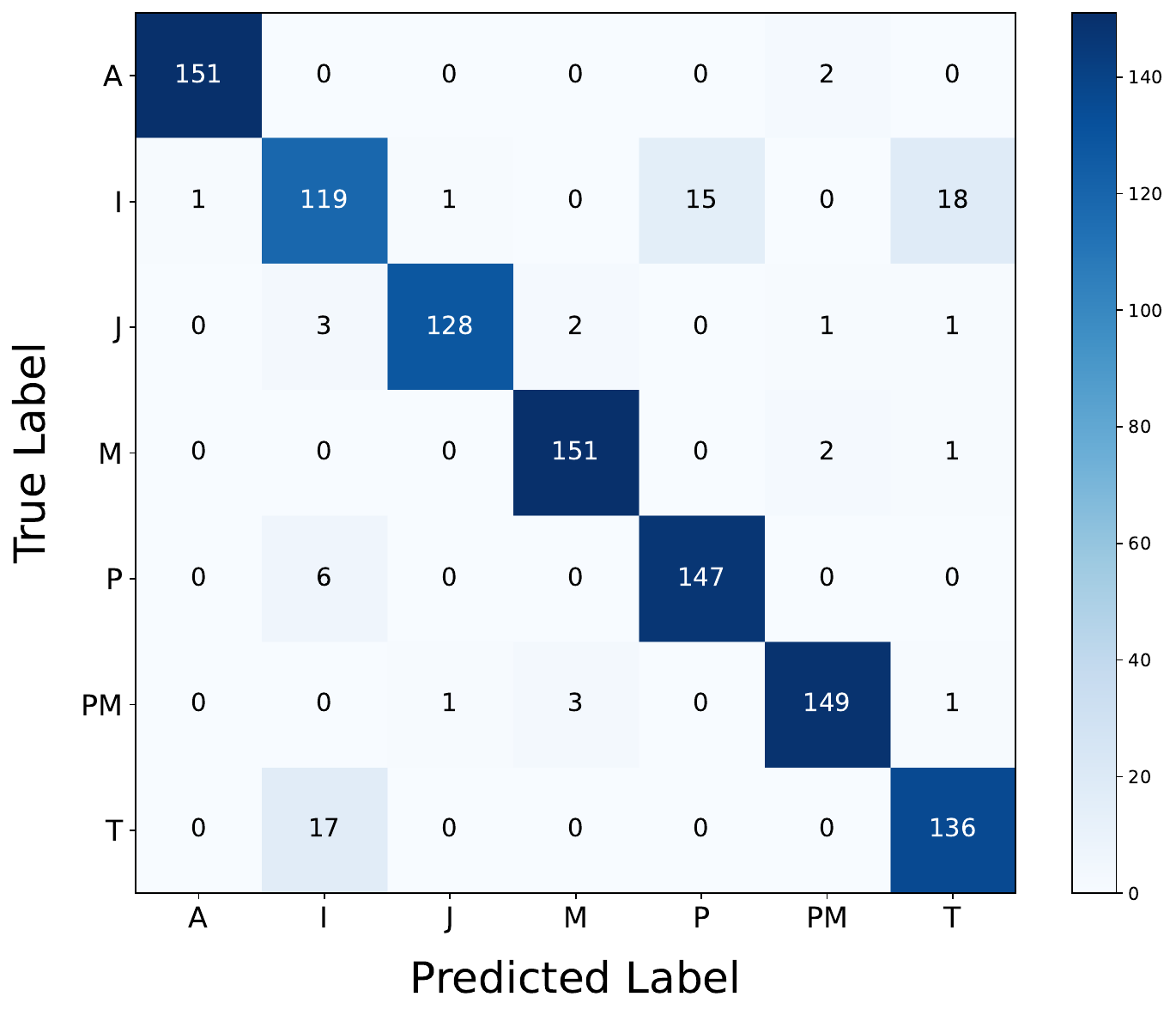} 
\caption{Confusion matrices of the SGAN-based transfer-learning model on the CellCognition dataset for the five-class (top) and eight-class cell-cycle (middle), and on our homemade Zeiss dataset (bottom).}
\label{fig:conf_transfer_5_8_class}
\end{figure}
\clearpage

\section{Confusion matrices Transfer Learning on Nagao dataset}

\begin{figure}[!h]
\centering
\begin{subfigure}{0.4\textwidth}
    \centering
    \includegraphics[width=0.8\textwidth]{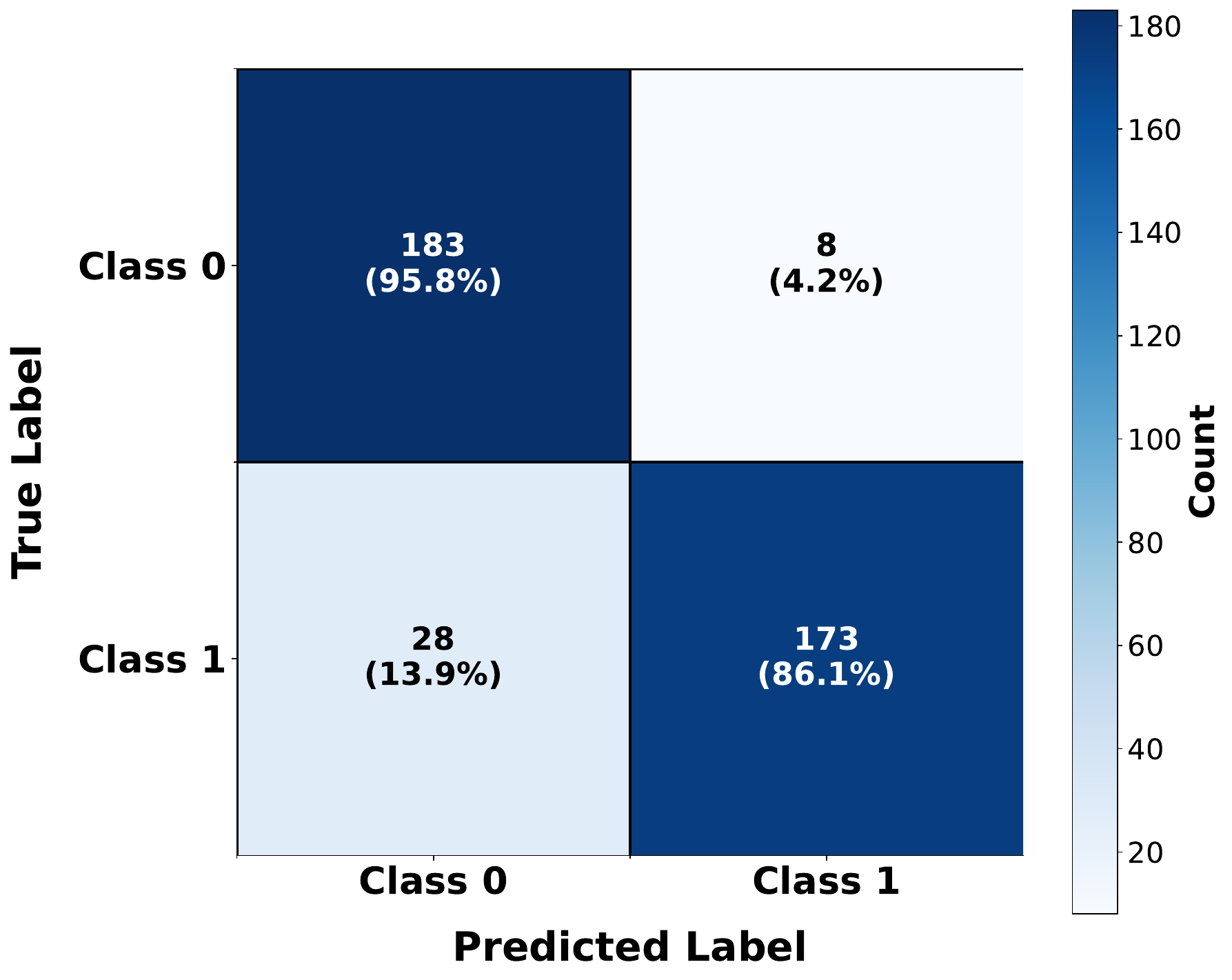}
    \caption{\textbf{B} HeLa-GM130\\$89.6 \pm 0.75\%$}
\end{subfigure}
\begin{subfigure}{0.4\textwidth}
    \centering
    \includegraphics[width=0.8\textwidth]{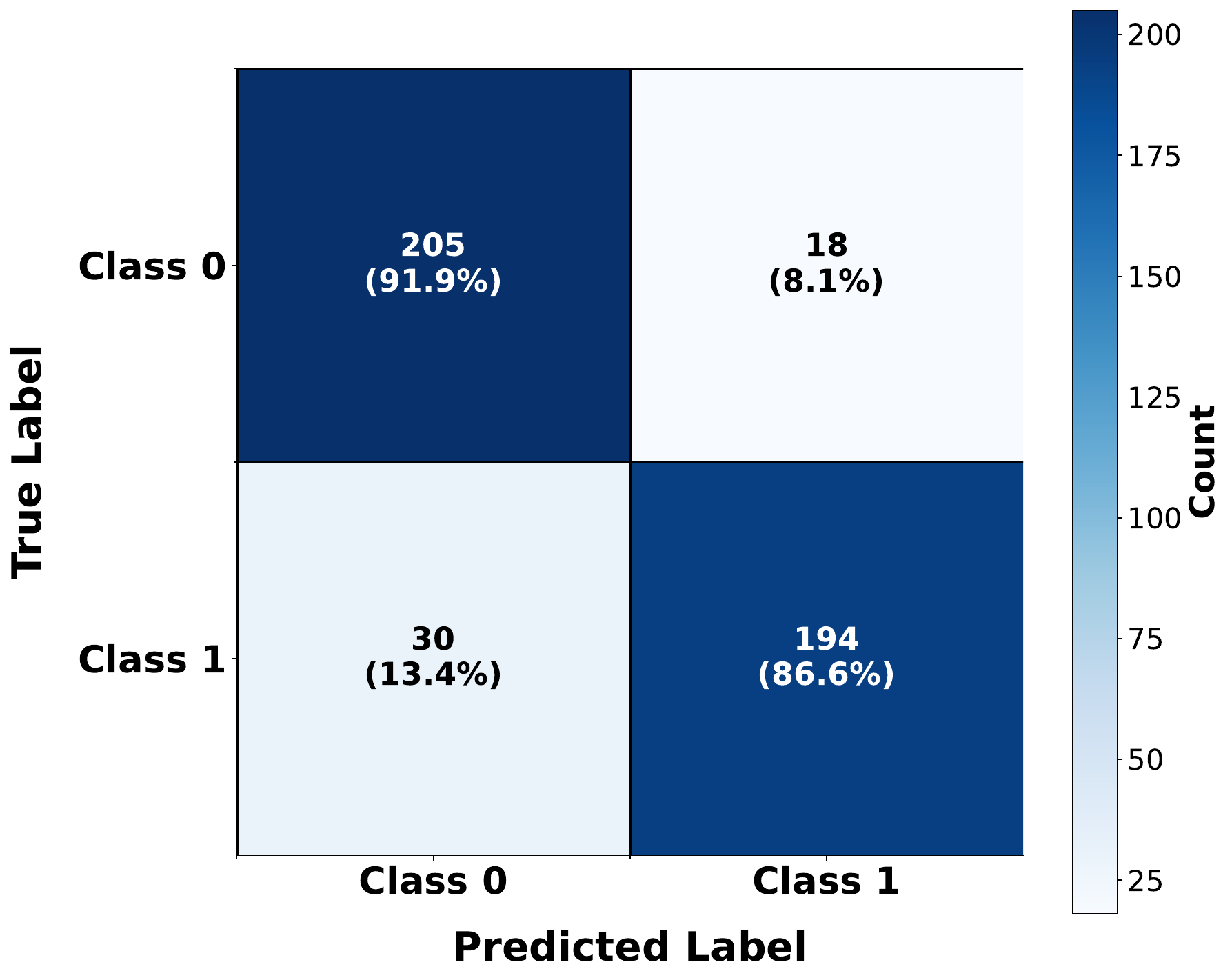}
    \caption{\textbf{C} NIH3T3\\$88.45 \pm 0.71\%$}
\end{subfigure}

\begin{subfigure}{0.4\textwidth}
    \centering
    \includegraphics[width=0.8\textwidth]{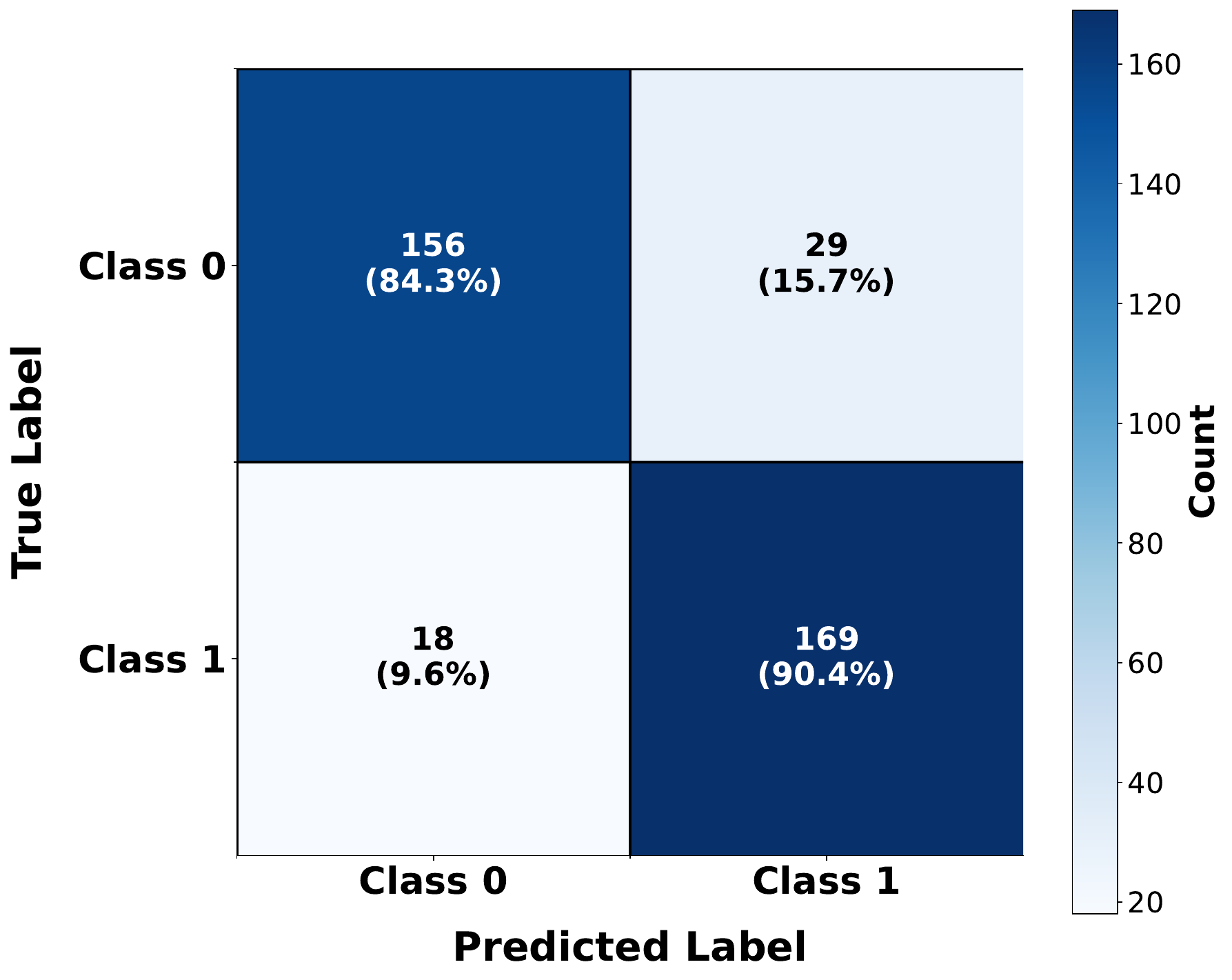}
    \caption{\textbf{D} RPE1-CENPF\\$86.65 \pm 0.86\%$}
\end{subfigure}
\begin{subfigure}{0.4\textwidth}
    \centering
    \includegraphics[width=0.8\textwidth]{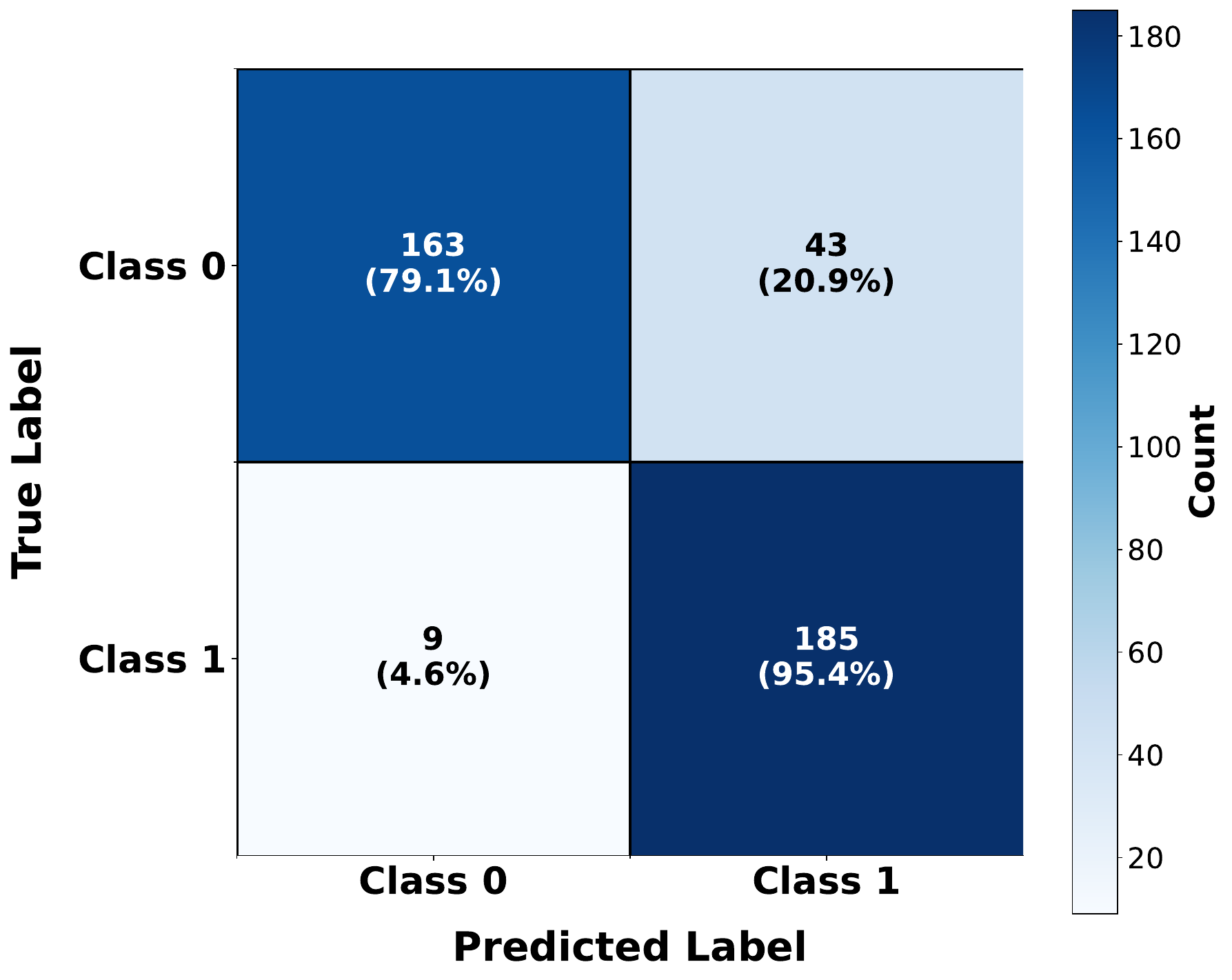}
    \caption{\textbf{E} HeLa-EB1\\$81.09 \pm 4.91\%$}
\end{subfigure}

\caption{Confusion matrices of transfer learning results across five G2-phase classification datasets.}
\label{fig:all_cm}
\end{figure}
\clearpage

\section{Confusion matrix Transfer Learning on DIC dataset}

\begin{figure}[!h]
\centering
\includegraphics[width=0.8\linewidth]{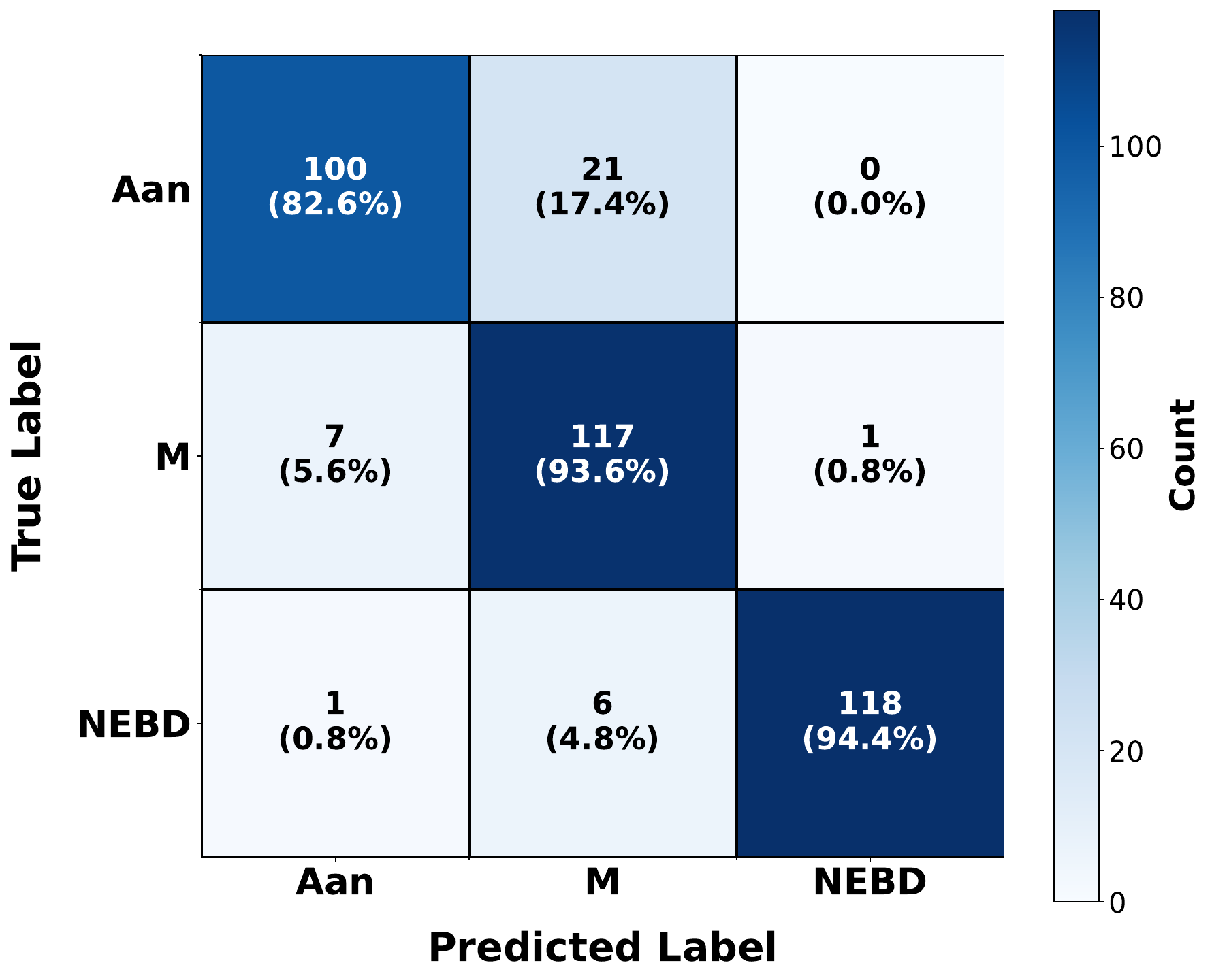}
\caption{Confusion matrix for the three-class classification on the DIC \textit{C.~elegans} embryo dataset.
Full retraining initialised from Mitocheck pre-trained weights substantially improved generalisation, achieving $84.13 \pm 7.42\%$ overall test accuracy.}
\label{figure:conf_mat_dic}
\end{figure}
\clearpage

\section{Transfer learning dataset}

\begin{table}[!h]
\centering
\small{
\begin{tabular}{lrrrr}
\toprule
\textbf{Dataset} & \textbf{Classes} & \textbf{Orig.} & \textbf{Aug.} & \textbf{Total} \\
\midrule
Cell Cog$\_5$ & 5 & 284 & 796 & 1,080 \\
Cell Cog$\_8$ & 8 & 2,024 & 7,244 & 9,268 \\
Homemade Zeiss & 7 & 1,321 & 3,963 & 5,284 \\
HHE & 2 & 501 & 1,503 & 2,004 \\
HHG & 2 & 491 & 1,470 & 1,961 \\
RHC & 2 & 461 & 1,396 & 1,857 \\
Cilia & 2 & 558 & 1,676 & 2,234 \\
DIC & 3 & 169 & 1,690 & 1,859 \\
\midrule
\textbf{Total} & & \textbf{5,809} & \textbf{19,738} & \textbf{25,547} \\
\bottomrule
\end{tabular}
}
\caption{Summary of image counts (original and augmented) across all datasets used in transfer learning experiments.}
\label{tab:dataset_image_counts}
\end{table}

\begin{figure}[!h]
  \centering
  \includegraphics[width=\linewidth]{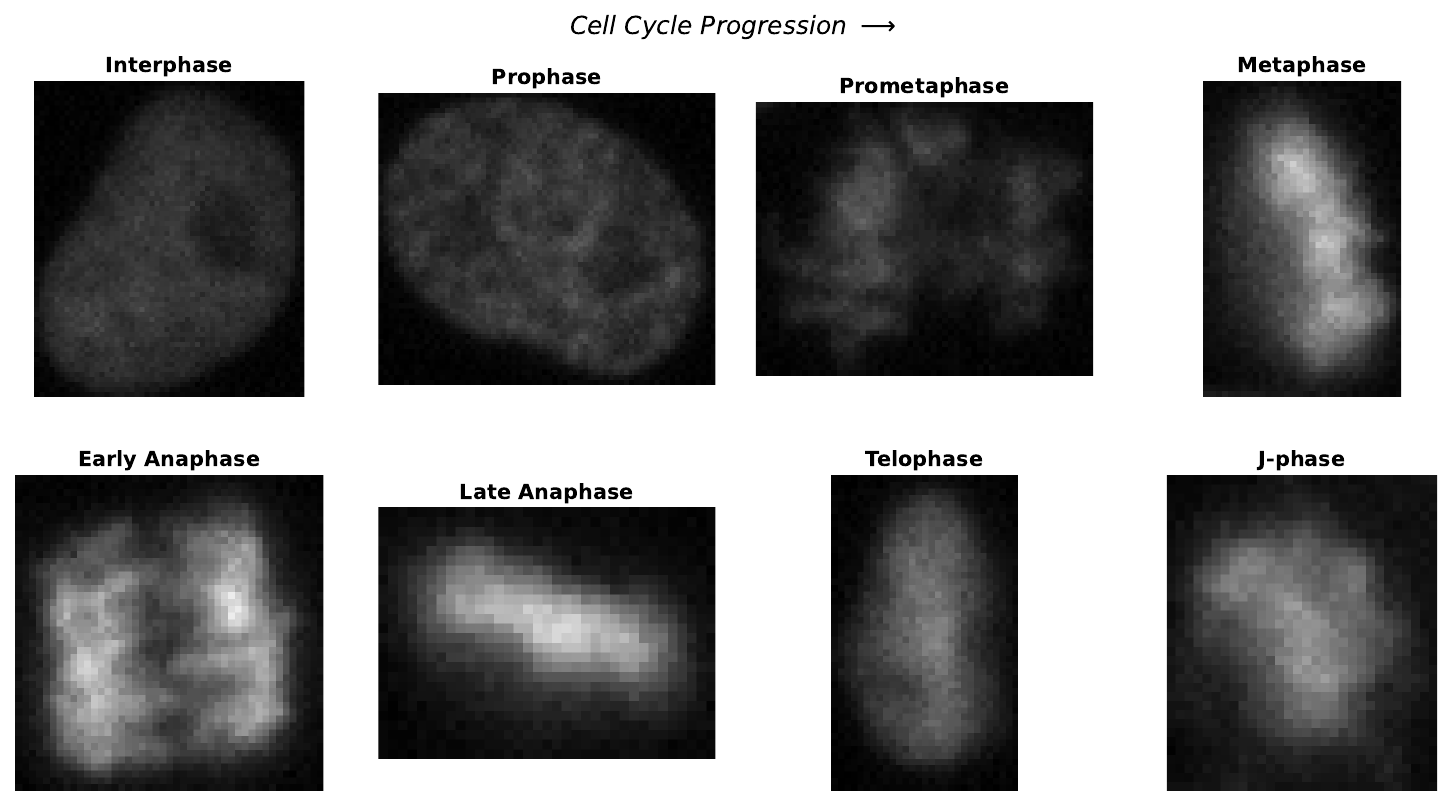}
  \caption{Representative sample image for each of the eight cell-cycle
           phases in the CellCognition dataset, displayed in order of
           mitotic progression: Interphase $\rightarrow$ Prophase
           $\rightarrow$ Prometaphase $\rightarrow$ Metaphase
           $\rightarrow$ Early Anaphase $\rightarrow$ Late Anaphase
           $\rightarrow$ Telophase $\rightarrow$ J-phase (cytokinesis).}
  \label{fig:cell_cycle_samples_cog}
\end{figure}

\begin{figure}[!h]
\centering

\begin{minipage}{0.1\textwidth}
    \centering
    \includegraphics[width=\linewidth]{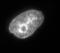}
\end{minipage}
\begin{minipage}{0.09\textwidth}
    \centering
    \includegraphics[width=\linewidth]{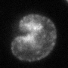}
\end{minipage}
\begin{minipage}{0.09\textwidth}
    \centering
    \includegraphics[width=\linewidth]{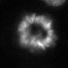}
\end{minipage}
\begin{minipage}{0.09\textwidth}
    \centering
    \includegraphics[width=\linewidth]{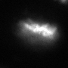}
\end{minipage}
\begin{minipage}{0.09\textwidth}
    \centering
    \includegraphics[width=\linewidth]{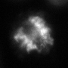}
\end{minipage}
\begin{minipage}{0.09\textwidth}
    \centering
    \includegraphics[width=\linewidth]{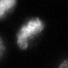}
\end{minipage}
\begin{minipage}{0.09\textwidth}
    \centering
    \includegraphics[width=\linewidth]{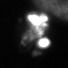}
\end{minipage}

\caption{Images acquired on a Zeiss inverted axio observer using a 20x dry objective.}

\label{figure:homemade}
\end{figure}

\begin{figure}[!h]
\centering
\footnotesize

\begin{tabular}{c c c}
\textbf{Dataset} & \textbf{G2 class} & \textbf{Not G2} \\[4pt]

\textbf{RPE1\_Hoechst\_CENPF} &
\includegraphics[width=0.23\linewidth]{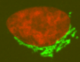} &
\includegraphics[width=0.23\linewidth]{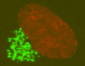} \\[10pt]

\textbf{HeLa\_Hoechst\_GM130} &
\includegraphics[width=0.23\linewidth]{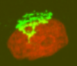} &
\includegraphics[width=0.23\linewidth]{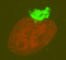} \\[10pt]

\textbf{HeLa\_Hoechst\_EB1} &
\includegraphics[width=0.23\linewidth]{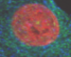} &
\includegraphics[width=0.23\linewidth]{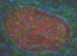} \\

\end{tabular}

\caption{Representative images from three G2-focused datasets: (top) RPE1\_Hoechst\_CENPF (RHC), (middle) HeLa\_Hoechst\_GM130 (HHG), and (bottom) HeLa\_Hoechst\_EB1 (HHE). Each dataset shows cells in G2 phase alongside non-G2 cells, demonstrating morphological markers used for binary cell cycle classification.}
\label{fig:RHC_HHG_HHE}
\end{figure}

\begin{figure}[!h]

\centering
\begin{minipage}{0.3\textwidth}
    \centering
    \includegraphics[width=\linewidth]{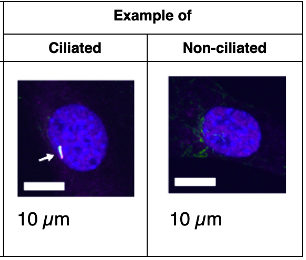}
\end{minipage}
\caption{Example images from ciliated / non-ciliated cells dataset.}
\label{fig:ciliated}
\end{figure}

\begin{figure}[!h]
\centering
\begin{minipage}{0.3\textwidth}
    \centering
    \includegraphics[width=\linewidth]{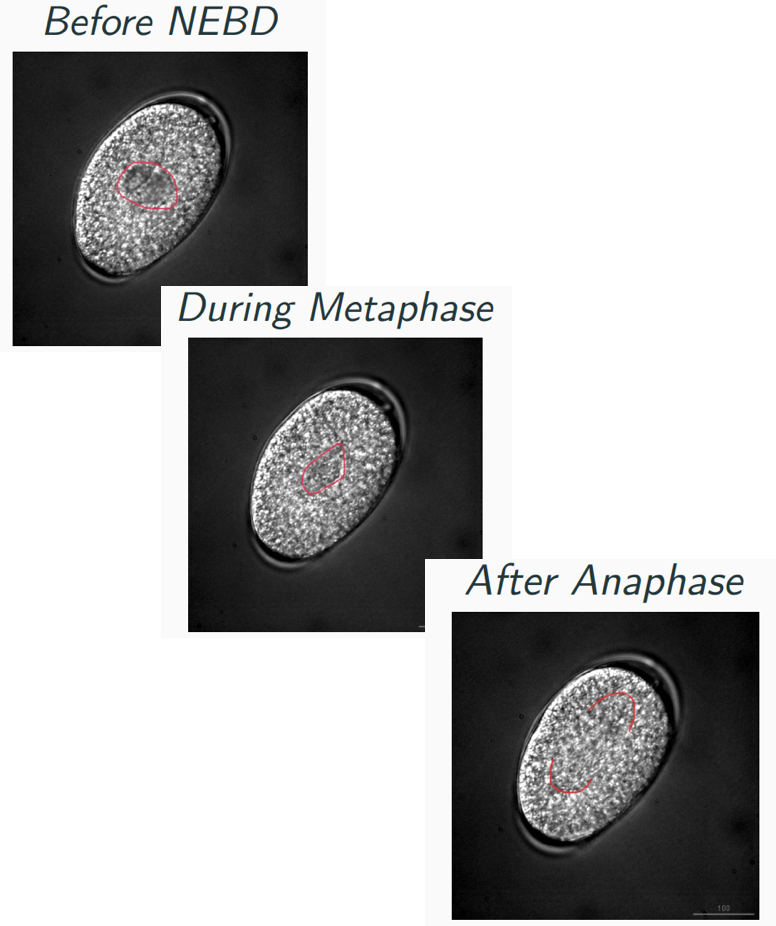}
\end{minipage}
\caption{Representative images from the C. elegans DIC Nomarski dataset showing three key cell cycle phases: before nuclear envelope breakdown (NEBD), during metaphase, and after anaphase. The dataset comprises 169 images.}
\label{figure:dic}
\end{figure}

\end{document}